\documentclass[mathematics,article,submit,pdftex,moreauthors]{Definitions/mdpi} 

\firstpage{1} 
\pubvolume{1}
\issuenum{1}
\articlenumber{0}
\pubyear{2024}
\copyrightyear{2024}
\datereceived{ } 
\daterevised{ }
\dateaccepted{ } 
\datepublished{ } 

\hreflink{https://doi.org/}

\usepackage{amsmath}
\usepackage{caption} 
\usepackage{subcaption}
\usepackage{textgreek}

\Title{Disentangling sources of multifractality in time series}

\TitleCitation{Title}

\Author{Robert Kluszczyński $^{1,2}$\orcidA{}, Stanisław Drożdż $^{1,3}$\orcidB{}, Jarosław Kwapień $^{1}$\orcidC{}, Tomasz Stanisz $^{1}$\orcidD{}, and Marcin Wątorek $^{3,4}$\orcidE{}}

\AuthorNames{Robert Kluszczyński, Stanisław Drożdż, Jarosław Kwapień, Tomasz Stanisz and Marcin Wątorek}

\AuthorCitation{Kluszczyński, R.; Drożdż, S.; Kwapień, J.; Stanisz, T.; Wątorek, M.}

\address{%
$^{1}$ \quad Complex Systems Theory Department, Institute of Nuclear Physics, Polish Academy of Sciences, ul. Radzikowskiego 152, \mbox{31-342 Krak\'ow, Poland;}\\
$^{2}$ \quad Faculty of Mathematics and Computer Science, Jagiellonian University, ul. Łojasiewicza 6, \mbox{30-348 Kraków, Poland;}\\
$^{3}$ \quad Faculty of Computer Science and Telecommunications, Cracow University of Technology, \mbox{31-155 Krak\'ow, Poland;} \\
$^{4}$ \quad Adapt Centre, School of Computing, Dublin City University, D02 PN40 Dublin, Ireland
}
\corres{Correspondence: stanislaw.drozdz@ifj.edu.pl (S.D.)}

\abstract{This contribution addresses the question commonly asked in scientific literature about the sources of multifractality in time series. Two primary sources are typically considered. These are temporal correlations and heavy tails in the distribution of fluctuations. Most often, they are treated as two independent components, while true multifractality cannot occur without temporal correlations. The distributions of fluctuations affect the span of the multifractal spectrum only when correlations are present. These issues are illustrated here using series generated by several model mathematical cascades, which by design build correlations into these series. The thickness of the tails of fluctuations in such series is then governed by an appropriate procedure of adjusting them to $q$-Gaussian distributions, and $q$ is treated as a variable parameter that, while preserving correlations, allows to tune these distributions to the desired functional form. Multifractal detrended fluctuation analysis (MFDFA), as the most commonly used practical method for quantifying multifractality, is then used to identify the influence of the thickness of the fluctuation tails in the presence of temporal correlations on the width of multifractal spectra. The obtained results point to the Gaussian distribution, so $q=1$, as the appropriate reference distribution to evaluate the contribution of fatter tails to the width of multifractal spectra. An appropriate procedure is presented to make such estimates.}


\begin{document}

\section{Introduction}
\label{sect::introduction}

Multifractality is an extremely useful concept for quantifying rough structures ubiquitous in nature~\cite{StanleyHE-1998a,BarabasiAL-1991a}. The most common use of this concept is to describe the patterns of non-differentiable time series in a unified way~\cite{BacryE-2001a,KantelhardtJ-2011a,SalatH-2017a,JiangZQ-2019a}. Such time series often exhibit patterns that differ across time scales and range of fluctuations. Multifractal analysis provides a way to model such a complex dynamics by identifying different scaling behaviors in high-intensity vs. low-intensity periods, by enabling better understanding of the data structure and a potential for extreme events~\cite{AusloosM-2002a,IhlenEAF-2013a,DrozdzS-2016a,TakaishiT-2018a,JiangZQ-2019a,KlamutJ-2020a}. It has proven its usability in a wide range of scientific disciplines, as it has been applied to study data originating in such diverse systems as financial markets~\cite{JiangZQ-2019a}, human brain~\cite{WatorekM-2024a}, natural language~\cite{StaniszT-2024a}, and climate~\cite{KantelhardtJ-2006a}. Measuring multifractality in time series involves analyzing how different moments or fluctuations scale across various time scales~\cite{ChhabraAB-1989a,ArneodoA-1995a,KantelhardtJ-2002a}. The key is to assess the singularity spectrum $f(\alpha)$, which reflects a range of fractal dimensions that describe the structure and distribution of singularities determined by the local scaling of fluctuations and quantified in terms of by the H\"older exponents $\alpha$~\cite{HalseyTC-1986a}. There are two driving factors that determine the values of $\alpha$ and their distribution: (1) temporal correlations are essential, because they set patterns of changes in adjacent values in a series of data and if such correlations exist, then (2) the distribution of fluctuations also matters. It should be expected that the greater range of fluctuations in the sense of the presence of large events will result in a greater range of the variability of $\alpha$. The shape of $f(\alpha)$ reflects, thus, the internal proportions in the organization of data in the time series, which originate from the long-range temporal correlations. 

In the light of the above, the questions that are often asked in scientific literature~\cite{IvanovPC-1999a,KantelhardtJ-2002a,MatiaK-2003a,KwapienJ-2005b,BarunikJ-2012a,GomesLF-2023a} about what is the source of multifractality in a given situation $-$ time correlations or fat tails of the fluctuation distribution $-$ as if these were two independent factors, are rather unfounded. The standard procedure of destroying correlations by shuffling the original time series~\cite{SchreiberT-1996a,TheilerJD-1996a} can often give an impression of obtaining a series that is characterized by a spectrum $f(\alpha)$ of the multifractal type, but actually, these are artifacts of the finiteness of the series as systematic studies show~\cite{DrozdzS-2009a,ZhouWX-2012a,KwapienJ-2023a}. On the one hand, for a multifractal time series whose probability distribution function is not stable in terms of the L\'evy criterion, the fat tails can only affect the width of $f(\alpha)$ if correlations are present~\cite{DrozdzS-2009a}. For the L\'evy-flight-type time series, on the other hand, the apparent multifractality of $f(\alpha)$ is a numerical artifact stemming from bifractality that cannot be properly grasped if the time series is too short~\cite{NakaoH-2000a,DrozdzS-2009a,RakR-2018a}. In the contribution presented here, a method is proposed to assess the impact of the distribution of fluctuations on the width of the singularity spectrum $f(\alpha)$ with a given degree of correlation. The presented methodology is based on synthetic time series generated from several types of multiplicative cascades~\cite{BarralJ-2015a}. By modifying the fluctuation PDFs for a specific cascade, so that they are described by $q$-Gaussian distributions with varying $q$ instead of the cascade's original PDF, the variability of $f(\alpha)$ as a function of $q$ is examined and the lower limit of its width is figured out. Possible surpluses relative to this lower bound can thus be naturally interpreted as being generated by fatter tails.

There are several practical algorithms for analyzing multifractal characteristics of time series. Among them, the most widely used $-$ due to its stability $-$ is multifractal detrended fluctuation Analysis (MFDFA)~\cite{KantelhardtJ-2002a,OswiecimkaP-2006a}. It was designed to deal with non-stationary data and effectively quantify multifractality by analyzing how fluctuations scale with the length of the segment. The above-mentioned analyzes will therefore be carried out by using this method.

\section{Multifractal detrended fluctuation analysis}
\label{sect::mfdfa}

\noindent
The essential steps of the MFDFA algorithm go as follows:

\vspace{0.1cm}
\noindent
\textit{Step 1.} A time series $U=\{u_i\}_{i=1}^T$ of $T$ consecutive measurements of some observable $u$ is partitioned into $M_s$ non-overlapping windows of length $s$ starting from both ends of $U$. This results in $2 M_s$ such windows.

\noindent
\textit{Step 2.} Possible non-stationarity of the signal in each window is eliminated by applying a detrending procedure to an integrated signal $X=\{X_i\}_{i=1}^s$, also termed a signal profile, whose elements read
\begin{equation}
X_i = \sum_{j=1}^i u_j.
\end{equation}
One of the possible ways of detrending $X$ is to use a polynomial $P^{(m)}$ of order $m$ that in each window $\nu=0,\ldots,2 M_s-1$ provides the best-fit to $X$. The higher is $m$, the better is the trend removal, but one has to be careful here in order not to overfit the data. Throughout this study, $m=2$ will be used, which seems to be an optimal choice~\cite{OswiecimkaP-2006a,OswiecimkaP-2013a}.

\noindent
\textit{Step 3.} The segment-wise variance of the residual detrended signal is calculated:
\begin{equation}
f^2(\nu,s) = {1 \over s} \sum_{i=1}^s (x_i - P^{(m)}(i))^2
\label{eq::variance}
\end{equation}
for all the segments $\nu$.

\noindent
\textit{Step 4.} A family of fluctuation functions of order $r$ ($r \in \mathbb{R}$) is defined on the average variance:
\begin{equation}
F_r(s) = \left\{ {1 \over 2M_s} \sum_{\nu=0}^{2M_s-1} \left[ f^2(\nu,s) \right]^{r/2} \right\}^{1/r}.
\label{eq::fluctuation.functions}
\end{equation}
In a certain range of positive and negative values of the parameter $r$ for which the moments are well defined, the functions $F_r(s)$ are then calculated for different values of scale $s$. Typically, the lower limit of $s$ is chosen above the length of the longest sequence of constant values of $U$ (in order to avoid problems with the summed terms in Eq.~(\ref{eq::fluctuation.functions}) for $r < 0$) and the upper limit of $s$ is $T/5$ (in order to sum over at least 10 segments in Eq.~(\ref{eq::fluctuation.functions})). The index $r$ can be viewed as related to the moments of the signal, so that its extreme values cannot be too large for a time series with heavy-tailed PDF of fluctuations.

\noindent
\textit{Step 5.} For a fractal signal, the fluctuation functions must depend on $s$ according to the power laws:
\begin{equation}
F_r(s) \sim s^{h(r)}.
\label{eq::scaling}
\end{equation}
A time series under study is considered multifractal when $h(r)$ depends on $r$ and monofractal otherwise. The function $h(r)$ is called the generalized Hurst exponent, because for $r=2$, $h(r)=H$, where $H$ is the standard Hurst exponent~\cite{HurstHE-1951a,HeneghanC-2000a}. The fractal properties of data manifest themselves if, for all considered values of $r$, $F_r(s)$ can be approximated by a straight line on a double-logarithmic plot.

\noindent
\textit{Step 6.} A conventional way of expressing the multifractality of data is the singularity spectrum $f(\alpha)$. It is derived from $h(r)$ by using the Legendre transform:
\begin{eqnarray}
\nonumber
\alpha = h(r) + rh'(r),\\
f(\alpha) = r \left[\alpha-h(r)\right] + 1,
\label{eq::singularity.spectrum}
\end{eqnarray}
where $\alpha$ measures the strength of a local singularity and is equivalent to the H\"older exponent~\cite{HalseyTC-1986a}. In geometrical terms, the function $f(\alpha)$ can be interpreted as a fractal dimension of the subset of the entire data set with the H\"older exponent equal to $\alpha$. For a monofractal time series, the pair $(\alpha,f(\alpha))$ is a single point, while for a multifractal it assumes a concave shape with its shoulders pointing down. The broader the singularity spectrum is, the richer is the multifractality of a time series. It can thus be viewed as a measure of time series complexity. Quantitatively, it can be measured by the spectrum width:
\begin{equation}
\Delta\alpha := \alpha_{\rm max}-\alpha_{\rm min} = \alpha(r_{\rm min})-\alpha(r_{\rm max}).
\label{eq::falpha.width}
\end{equation}
Typically, $f(\alpha)$ is symmetric, but it also often happens that its shape is distorted and asymmetric, which indicates that the time series points of different magnitude exhibit different hierarchical organization~\mbox{\cite{OhashiK-2000a,CaoG-2013a,DrozdzS-2015a,GomezGomezJ-2021a}.} These characteristics may alternatively be expressed in terms of the multifractal spectrum $\tau(r)$ defined by
\begin{equation}
\tau(r) = rh(r)-1.
\end{equation}
For monofractal time series, $\tau(r)$ depends linearly on $r$ (because $h(r)$ is constant then), while it is nonlinear for multifractal ones.

\section{Multifractal properties of the multiplicative cascades}
\label{sect::multiplicative.cascades}

In this Section, a few examples of the multiplicative cascades are considered. For all these cascades, exact analytical results are available making them convenient models for a controlled use of MFDFA. It is worthwhile to point out that the multiplicative cascades are long-range autocorrelated by construction and these correlations are both linear and nonlinear. This means that the multifractal properties of the cascades are genuine and cannot be considered as a numerical artifact.

In more intuitive terms the mathematical cascades refer to processes where a small change or event triggers a chain reaction, leading to significant outcomes. These cascades occur in systems where one event influences others in a sequence, like dominoes falling.

\subsection{Deterministic binomial cascade}

\begin{figure}
\includegraphics[width=14 cm]{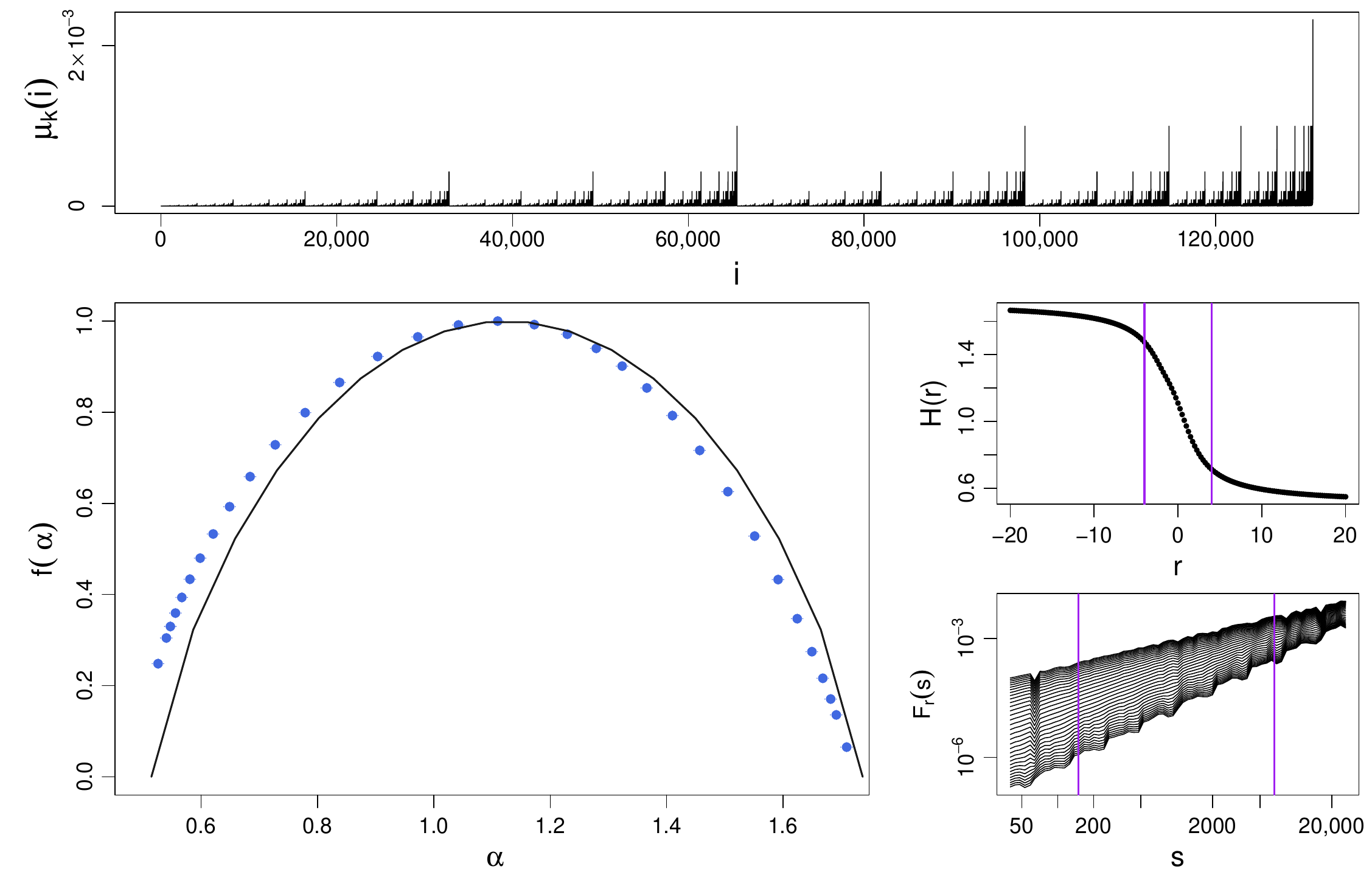}
\caption{(Top) Time series of a deterministic binomial multiplicative cascade with $p=0.3$ and $k=17$ iterations ($N=131,072$ data points). (Bottom left) Singularity spectrum $f(\alpha)$ for the cascade shown in the top panel calculated with MFDFA (blue symbols) and its theoretical form calculated from Eq.~(\ref{eq::binomial.cascade.falpha}) (black solid line). (Middle right) The generalized Hurst exponents $h(r)$ and  (bottom right) the fluctuations functions $F_r(s)$ are shown together with vertical lines denoting the range of $r$ ($-4 \le r \le 4$) and $s$ ($40 \le s \le N/5$), respectively, considered for the calculation of $f(\alpha)$. Each function: $f(\alpha)$, $h(r)$, and $F_r(s)$ has been averaged over 10 independent realizations of the process.}
\label{fig::binomial.cascade}
\end{figure}

The simplest deterministic cascade $-$ the binomial multiplicative cascade $-$ is well-known to be multifractal in nature~\cite{FederJ-1989a}. To construct it, one starts with the interval $[0,1]$, a uniformly distributed mass $\mu_0=1$ on this interval, and a multiplier $p\in(0,1)$. In the first step of the multiplicative process, the interval is divided into two subintervals of equal length $2^{-1}$ and the mass $\mu_1(0) = p\mu_0=p$ and $\mu_1(1) = (1-p)\mu_0=(1-p)$, respectively. The same is done to each subinterval in the second and subsequent steps, which produce $N=2^k$ subintervals in the $k$th step with the mass given by
\begin{equation}
\mu_k(i) = p^\varepsilon (1-p)^{k-\varepsilon},
\end{equation}
where $i=0,...,2^k-1$ labels the individual subintervals $[i/N,(i+1)/N]$. If $i/N = \overline{0,\eta_1,\eta_2,\dots,\eta_k}$ is the base-$2$ fractional representation with $\eta_j\in \{0,1\}$, then $\varepsilon = k-\sum_j \eta_j$ may denote the number of zeros in this representation. Since there are $\binom{k}{\varepsilon}$ numbers $i/N$ with exactly $\varepsilon$ zeros, the process conserves mass:
\begin{equation}
\sum_{i=0}^{N-1} \mu_k(i) = \sum_{\varepsilon=0}^k \binom{k}{\varepsilon}p^\varepsilon (1-p)^{k-\varepsilon} = (p+(1-p))^k=1
\end{equation}
and it is called \textit{microcanonical}, thus. Fig.~\ref{fig::binomial.cascade} (top) shows a realization of this process with $k=17$ iterations, which gives $N=2^{17}$ data points total. That this process develops a fat-tailed PDF, it is evident. It can be shown that the respective H\"older exponents and singularity spectrum are expressed by the following formulas~\cite{FederJ-1989a,KantelhardtJ-2002a}:
\begin{eqnarray}
\nonumber
\alpha(\varepsilon) = -\frac{\varepsilon \ln(p) + (k-\varepsilon) \ln(1-p)}{k\ln(2)},\\
f(\varepsilon) = -\frac{\varepsilon\ln(\varepsilon) + (k-\varepsilon)\ln(k-\varepsilon) - k\ln(k)}{k\ln(2)}.
\label{eq::binomial.cascade.falpha}
\end{eqnarray}
The singularity spectrum for a cascade with $p=0.3$ calculated via MFDFA as well as the theoretical spectrum calculated using the above equations is plotted in Fig.~\ref{fig::binomial.cascade} (bottom left). The fat-tailed PDF of the time series demands restricting the range of $r$ considered for the calculation of $f(\alpha)$, so the range $-4 \le r \le 4$ is used. However, the finite size of the sample allows one to show $h(r)$ for a broader range of $r$ $-$ see Fig.~\ref{fig::binomial.cascade} (middle right panel) $-$ just to observe that the properly restricted range almost saturates the variability range of $h(r)$ and little information on the multifractal structure is lost by imposing the restriction mentioned above. The singularity spectra are broad with $\Delta \alpha > 1$, which is a signature of rich multifractality, and almost symmetric, which may indicate that the multifractal structure is homogeneous across the entire range of fluctuation amplitudes. One can observe that the results from MFDFA show a satisfactory agreement with theory (for a detailed discussion on this issue see Ref.~\cite{OswiecimkaP-2006a}). It is also noteworthy to look at the fluctuation functions $F_r(s)$, which exhibit an approximate power-law form for the whole range of the considered scales and all the values of $r$ (bottom right). Such a model behavior of $F_r(s)$ is usually associated with high reliability of the obtained results as experience demonstrates~\cite{KantelhardtJ-2002a,OswiecimkaP-2006a,DrozdzS-2009a,OswiecimkaP-2013a}.


\subsection{Stochastic cascades}

The classical deterministic cascade described above is an example of a more general construction that involves dividing the interval $[0,1]$ with uniform mass into $b$ equal subintervals and randomizing the mass allocating multipliers. At each step, $b^k$ independent and identically distributed multipliers $M(\eta_1,\eta_2,\dots,\eta_k)$ will be drawn from a given non-negative random probability distribution $\mathcal{M}$. Then the mass allocated to the $i$th interval ($0\leq i\leq b^k-1$) is equal to
\begin{equation}
\mu_k(i) = M(\eta_1)M(\eta_1,\eta_2)\dots M(\eta_1,\eta_2,\dots,\eta_k)
\end{equation}
where the interval is $[i/b^k,(i+1)/b^k]$ and $i/b^k=\overline{0,\eta_1\eta_2\dots\eta_k}$ is a fraction expressed in base $b$, so $\eta_i \in \{0,...,b-1\}$. In this process, mass is preserved only on average, so one expects $E[\mathcal{M}]=1/b$. The cascades that conserve mass in a statistical sense are called \textit{canonical}~\cite{MandelbrotBB-1989a}. They were reported to exhibit multifractality~\cite{MandelbrotBB-1989a,MandelbrotBB-1997a} and they found applications in financial modeling, among others~\cite{CalvetL-2002a}.

\subsubsection{Log-normal cascade}

\begin{figure}
\includegraphics[width=14 cm]{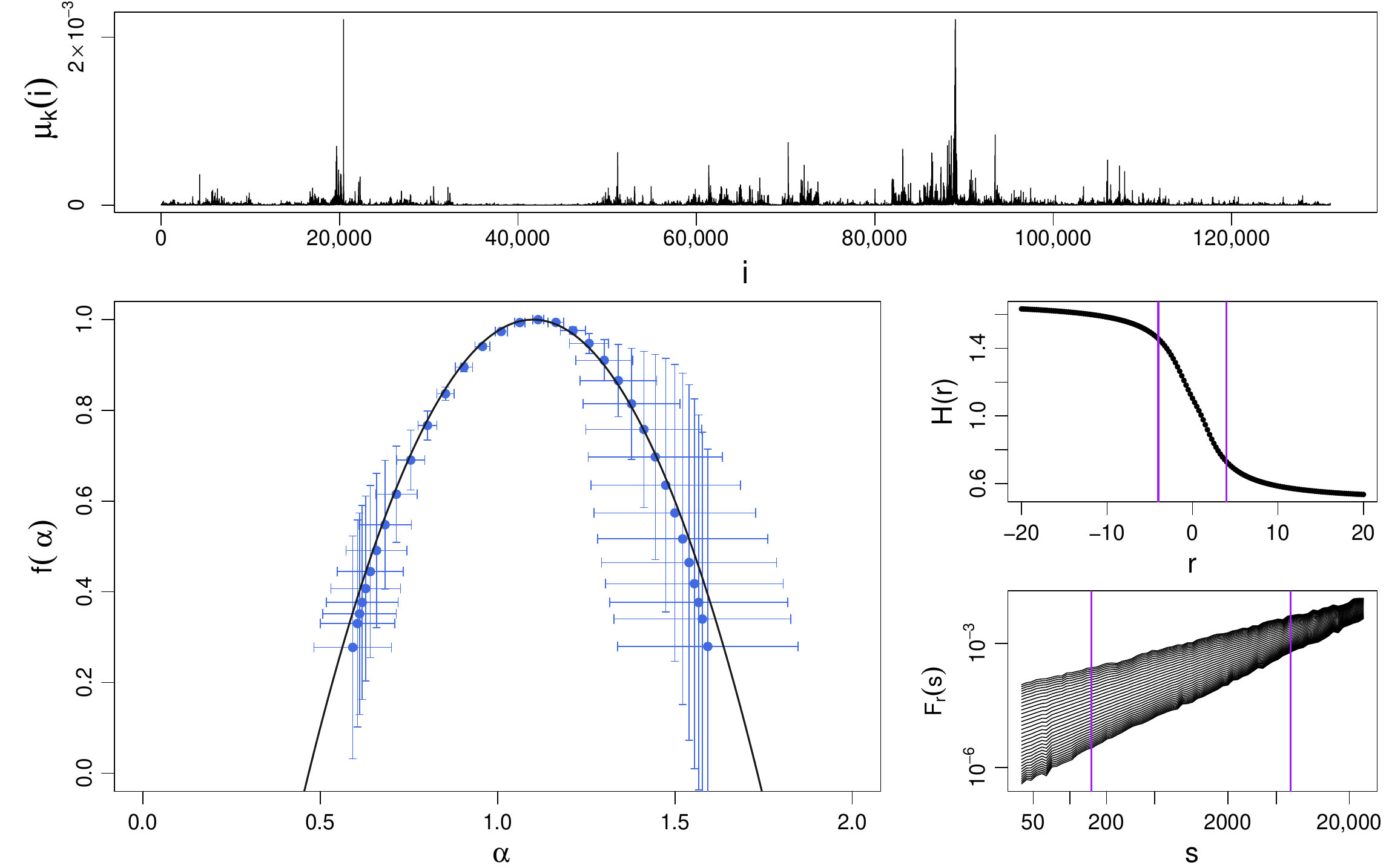}
\caption{(Top) Sample time series of a log-normal multiplicative cascade with $k=17$ iterations ($N=131,072$ data points), $\mu=1.1$, and $\sigma^2=1/5\ln2$. (Bottom left) Singularity spectrum for this cascade obtained by using MFDFA (blue symbols with error bars indicating standard deviation) and by using Eq.~(\ref{eq::log-normal.cascade.falpha}) (black solid line). (Middle right) the generalized Hurst exponents $h(r)$ and (bottom right) the fluctuations functions $F_r(s)$ are shown together with vertical lines denoting the range of $r$ and $s$, respectively, considered for the calculation of $f(\alpha)$. Each function: $f(\alpha)$, $h(r)$, and $F_r(s)$ was averaged over 10 independent realizations of the process.}
\label{fig::log-normal.cascade}
\end{figure}

The first example of a stochastic canonical cascade is the cascade whose multipliers $M(\eta_1,\eta_2,\dots,\eta_k)$ are taken from a log-normal distribution with PDF given by
\begin{equation}
p(x) = \cfrac{1}{x\sigma\sqrt{2\pi}} e^{-(\ln x-\mu)^2/2\sigma^2}, \quad x>0.
\label{eq::log-normal.pdf}
\end{equation}
In step $k$, the mass $\mu_k(i)$ of the $i$th interval reads
\begin{equation}
-\log_2(\mu_k(i)) = -\log_2\left(\prod_{j=1}^k M(\eta_1,\dots,\eta_j)\right) = \sum_{j=1}^k m(\eta_1,\dots,\eta_j),
\label{eq::log-normal.multipliers}
\end{equation}
where $m(\eta_1,\eta_2,\dots,\eta_k) := -\log_2(M(\eta_1,\eta_2,\dots,\eta_k))$ are taken from a normal distribution $N(\mu,\sigma)$. The sum on r.h.s. is likewise normally distributed with $N(k\mu,\sqrt{k}\sigma)$. Eq.~(\ref{eq::log-normal.multipliers}) allows one to develop a theoretical formula for the singularity spectrum $f(\alpha)$~\cite{CalvetL-1997a}, which has a parabolic shape:
\begin{equation}
f(\alpha) = 1 - \frac{1}{2 \ln{b}}\left(\frac{\alpha-\mu}{\sigma}\right)^2
\label{eq::log-normal.cascade.falpha}
\end{equation}
with maximum at $\alpha_0=\mu$. In order for this cascade to be canonical, the parameters $\mu$ and $\sigma^2$ need to satisfy~\cite{CalvetL-1997a}:
\begin{equation}
\mu = \frac{\sigma^2\ln{2}}{2} + \frac{\ln b}{\ln 2}
\label{eq::log-normal.cascade.canonical-condition}
\end{equation}
Fig.~\ref{fig::log-normal.cascade} shows a sample realization of this process (top panel) and the spectra $f(\alpha)$ from the MFDFA procedure and Eq.~(\ref{eq::log-normal.cascade.falpha}) (bottom left panel) for $b=2$, $\mu=1.1$, and $\sigma^2 = 1/5\ln{2}$. The shape of $f(\alpha)$ obtained with MFDFA confirms the theoretically predicted multifractal structure of the cascade ($\Delta\alpha \approx 1.2$) with almost a perfect agreement between the both within the error bars. In the present case, symmetry of $f(\alpha)$ is slightly distorted and its left shoulder seems to be longer than the right one. However, the significant error bars on the data points representing $r < 0$ (the right shoulder) make any judgment on whether the results from MFDFA exhibit an actual asymmetry rather impossible.


\subsubsection{Log-gamma cascade}

\begin{figure}
\includegraphics[width=14 cm]{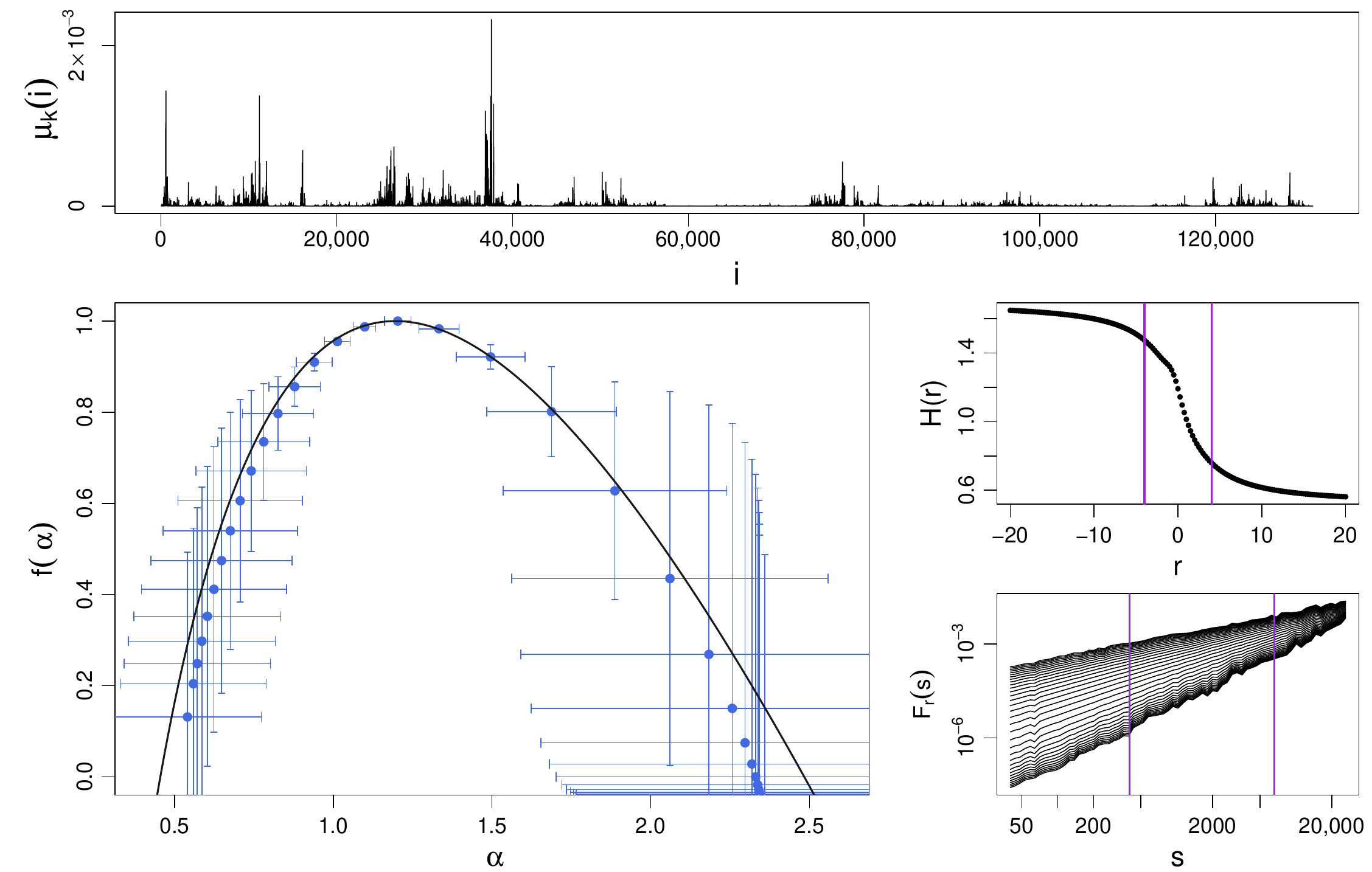}
\caption{(Top) Sample time series of a log-gamma multiplicative cascade with $k=17$ iterations ($N=131,072$ data points), and parameters $\gamma=2$ and $\beta=\ln2/(\sqrt{2}-1)$. (Bottom left) Singularity spectrum for this cascade obtained by using MFDFA (blue symbols with error bars indicating standard deviation) and by using Eq.~(\ref{eq::log-gamma.cascade.falpha}) (black solid line). (Middle right) the generalized Hurst exponents $h(r)$ and (bottom right) the fluctuations functions $F_r(s)$ are shown together with vertical lines denoting the considered range of $r$ and $s$, respectively, considered for the calculation of $f(\alpha)$. Each function: $f(\alpha)$, $h(r)$, and $F_r(s)$ was averaged over 10 independent realizations of the process.}
\label{fig::log-gamma.cascade}
\end{figure}

Another example of the distribution $\mathcal{M}$, from which the multipliers $M(\eta_1,\eta_2,\dots,\eta_k)$ are drawn, may be a log-gamma distribution $Gamma(\gamma,\beta)$:
\begin{equation}
p(x) =\frac{\beta^\gamma x^{-\beta-1} (\log x)^{\gamma-1}}{\Gamma(\gamma)},
\label{eq::log-gamma.cascade.pdf}
\end{equation}
where $\gamma,\beta>0$ are the shape and rate parameters, respectively. Like before, one can write the expression $-\log_2(\mu_k(i)) = \sum_{j=1}^k m(\eta_1,\eta_2,\dots,\eta_j)$ for the mass of an interval $i$. It is known that the sum of $k$ i.i.d. variables with PDFs given by $Gamma(\gamma,\beta)$ is also gamma-distributed with $Gamma(k\gamma,\beta)$. The theoretical singularity spectrum $f(\alpha)$ can be written as~\cite{CalvetL-1997a}:
\begin{equation}
f(\alpha) = 1+\gamma \log_b\left(\frac{\alpha\beta}{\gamma}\right)+\frac{\gamma-\alpha\beta}{\ln{b}},
\label{eq::log-gamma.cascade.falpha}
\end{equation}
whose maximum is reached for $\alpha_0=\gamma/\beta$. For this cascade to be canonical, $\gamma,\beta$ need to satisfy
\begin{equation}
1+\frac{\ln{2}}{\beta} = b^{1/\gamma}
\label{eq::log-gamma.cascade.canonical-condition}.
\end{equation}
Fig.~\ref{fig::log-gamma.cascade} shows a sample realization of the log-gamma process (top panel) and the spectra $f(\alpha)$ from MFDFA and Eq.~(\ref{eq::log-gamma.cascade.falpha}) (bottom left panel) for $b=2$, $\gamma=2$, $\beta=\ln 2 / (\sqrt{2}-1)$. Both the theoretical and MFDFA-derived singularity spectra are multifractal with $\Delta\alpha > 1.8$ and their asymmetry is unquestionable. It can be interpreted in such a way that the fluctuations with small amplitude (that correspond to $r < 0$ and the right shoulder) display a richer multifractality than the medium and large ones. The agreement between the data and theory is satisfactory here as well.


\subsubsection{Log-Poisson cascade}

\begin{figure}
\includegraphics[width=14 cm]{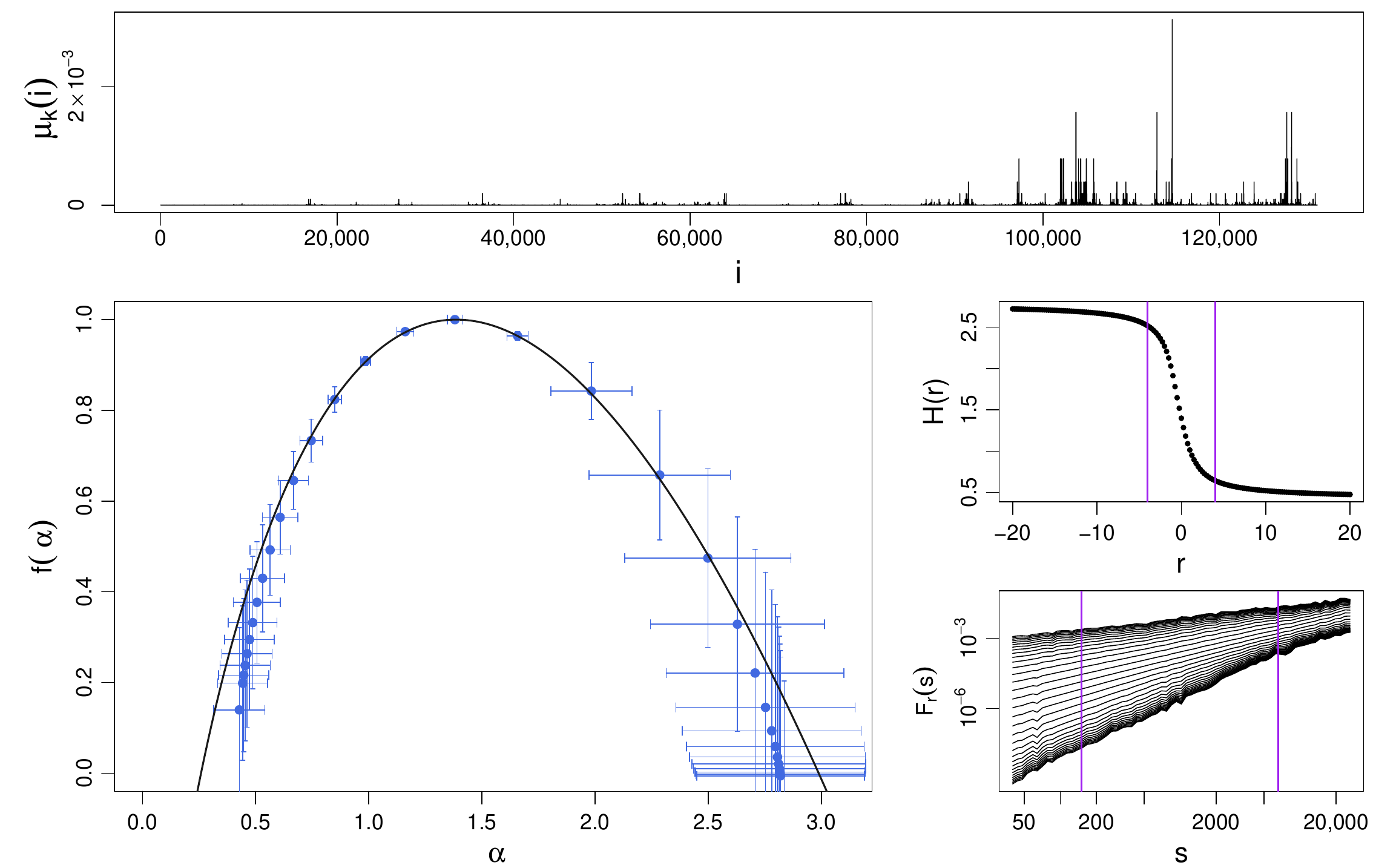}
\caption{(Top) Sample time series of a log-Poisson multiplicative cascade with $k=17$ iterations ($N=131,072$ data points), and parameter $\lambda=2 \ln 2$. (Bottom left) Singularity spectrum for this cascade obtained by using MFDFA (blue symbols with error bars indicating standard deviation) and by using Eq.~(\ref{eq::log-poisson.cascade.falpha}) (black solid line). (Middle right) the generalized Hurst exponents $h(r)$ and (bottom right) the fluctuations functions $F_r(s)$ are shown together with vertical lines denoting the considered range of $r$ and $s$, respectively, considered for the calculation of $f(\alpha)$. Each function: $f(\alpha)$, $h(r)$, and $F_r(s)$ was averaged over 10 independent realizations of the process.}
\label{fig::log-poisson.cascade}
\end{figure}

Finally, let one consider a discrete distribution, namely the Poisson distribution $Poisson(\lambda)$, whose PDF is given by
\begin{equation}
p(k) = \frac{\lambda^k e^{-\lambda}}{k!},
\label{eq::log-poisson.cascade.pdf}
\end{equation}
where $\lambda>0$ is the rate parameter. Its logarithmic version has the same form as above but for a new random variable $y=e^k$ instead of $x=k$. Given that the sum of $k$ i.i.d. random variables with the Poisson distribution is likewise Poisson-distributed with the parameter $k\lambda$, it is possible to derive the analytical form of the singularity spectrum~\cite{CalvetL-1997a}:
\begin{equation}
f(\alpha) = 1 - \frac{\lambda}{\ln{b}}+\alpha \log_b\left(\frac{\lambda e}{\alpha}\right),
\label{eq::log-poisson.cascade.falpha}
\end{equation}
which assumes its maximum value at $\alpha_0=\lambda$. For a log-Poisson cascade to be canonical, one must have
\begin{equation}
\lambda=2\ln b .
\label{eq::log-poisson.cascade.canonical-condition}
\end{equation}
Fig.~\ref{fig::log-poisson.cascade} shows a sample realization of the log-Poisson process (top panel) and the spectra $f(\alpha)$ from MFDFA and Eq.~(\ref{eq::log-poisson.cascade.falpha}) (bottom left panel) for $b=2$, $\lambda=2 \ln 2$. One can notice that the mass $\mu_k(i)$ has quantized values because the distribution is discrete. As for the log-gamma cascade above, here $f(\alpha)$ is also broad ($\Delta\alpha > 1.8$) and right-side asymmetric, which is shown by both the data and theory.


\section{$q$-Gaussian distribution}
\label{sect::qgaussians}

\begin{figure}
\includegraphics[width=14 cm]{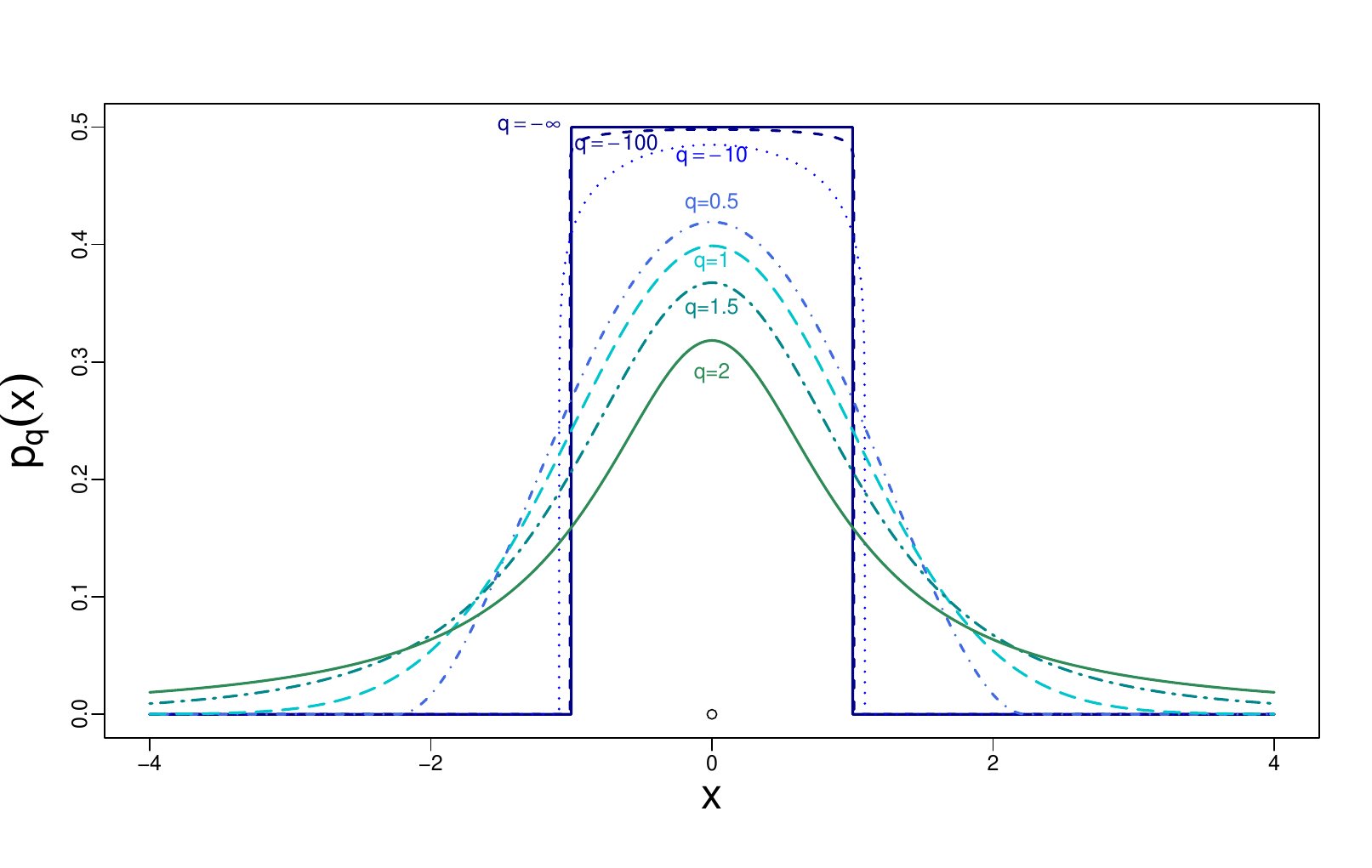}
\caption{Probability density functions of the $q$-Gaussian distributions $p_q(x)$ for selected values of the Tsallis parameter: $q=-\infty,-100,-10,0.5,1,1.5,2$. For $q=1.5$ and $q=2$, one can observe increasingly fat tails. For $q=1$, the normal distribution is restored. The finite support of $p_q(s)$ becomes visible for $q=0.5$ and the smaller values of $q$.}
\label{fig::qgaussian.pdf}
\end{figure}

The family of $q$-Gaussian distributions has seen a variety of applications in statistical physics and in modeling of empirical data sets from the measurements of observables in natural systems. Statistical mechanics~\cite{TsallisC-2009a,LutzE-2003a,TsallisC-2009b}, astrophysics~\cite{BurlagaLF-2005a,BurlagaLF-2006a,BurlagaLF-2007a}, geology\cite{CarusoF-2007a}, financial modeling and economics~\cite{BorlandL-2002a,RakR-2007a,DrozdzS-2010a,WatorekM-2021a} are only a few examples of the broad spectrum of the successful applications of $q$-Gaussian distributions. These distributions may be viewed as a generalization of the Gaussian distribution, in the same way as the Tsallis entropy $S_q$ generalizes the Boltzmann-Gibbs entropy $S$~\cite{UmarovS-2008a,TsallisC-1995a}. The $q$-Gaussian distribution $\mathcal{G}_{q}$ has two parameters: the shape parameter $q\in(-\infty,3)$ and the width parameter $\beta > 0$. Its PDF has the following form~\cite{UmarovS-2008a}:
\begin{equation}
p_q(x) = \frac{\sqrt{\beta}}{C_q} e_q(-\beta x^2),
\label{eq::qgaussian.pdf}
\end{equation}
where $e_q(x)$ is the $q$-$exponential$ function defined by
\begin{equation}
e_q(x) = \begin{cases}
    (1+(1-q)x)^{1/(1-q)} & \text{if } q\neq 1 \text{ and } 1+(1-q)x>0\\
    0 & \text{if } q\neq 1 \text{ and } 1+(1-q)x\leq 0\\
    e^x & \text{if } q=1
\end{cases}
\end{equation}
and $C_q=\int_{-\infty}^{\infty} e_q(x^2)dx$ is a normalization factor. It can be shown~\cite{UmarovS-2008a} that
\begin{equation}
C_q = \begin{cases}
    \frac{\sqrt{\pi}\Gamma\left(\frac{3-q}{2(q-1)}\right)}{\sqrt{q-1}\Gamma\left(\frac{1}{q-1}\right)} & \text{for } 1<q<3\\
    \sqrt{\pi} & \text{for } q=1\\
    \frac{2\sqrt{\pi}\Gamma\left(\frac{1}{1-q)}\right)}{(3-q)\sqrt{1-q}\Gamma\left(\frac{3-q}{2(1-q)}\right)} & \text{for } -\infty <q<1 .
\end{cases}
\end{equation}
From a perspective of the present study, random variables with the $q$-Gaussian PDF are especially useful, because they can model a variety of processes with different shapes of their PDFs depending on a value of the parameter $q$: uniform on a compact support ($q \to -\infty$), unimodal on a compact support ($-\infty < q < 1$), Gaussian ($q=1$), leptokurtic with power-law tails ($1 < q < 3$), and uniform flat on infinite support ($q \to 3$). The distribution cannot be normalized for $q \geqslant 3$. Examples of the distributions $\mathcal{G}_q$ for a few selected values of $q$ are shown in Fig.~\ref{fig::qgaussian.pdf}. Variance of the $q$-Gaussian PDFs
\begin{equation}
\sigma^2(p_q) = \frac{1}{\beta(5-3q)}
\end{equation}
is finite for $q < 5/3$ and infinite for $q \geqslant 5/3$. This means that, under no-correlation condition, the $q$-Gaussians reside in the (standard) Gaussian basin of attraction in the former case (the central limit theorem) and in the L\'evy basin of attraction in the latter case (the L\'evy-Gnedenko limit theorem)~\cite{TsallisC-2009b}. However, because of increasingly fat tails, convolutions of the $q$-Gaussians may tend to the limit distribution very slowly~\cite{PratoD-1999a,DrozdzS-2009a}. There is a simple relation between $q$ and the L\'evy parameter $\alpha_{\rm L}$ of the limit stable distribution~\cite{PratoD-1999a}:
\begin{equation}
\alpha_L = (3-q)/(q-1)
\label{eq::levy.qgaussian}
\end{equation}
valid for $5/3 \leqslant q < 3$. When $q \to 1$, the normal distribution is retrieved:
\begin{equation}
p_1(x) = \frac{1}{\sqrt{\pi/\beta}} e^{-\beta x^2} = N\left(0,\sigma = \tfrac{1}{2\beta}\right).
\label{eq:qgaussian.normal}
\end{equation}
For $q < 1$, the $q$-Gaussian distribution becomes bounded with a compact support given by
\begin{equation}
\mathrm{supp}(p_q) =  \left[ -\frac{1}{(1-q)\beta},\frac{1}{(1-q)\beta}\right].
\end{equation}

As we shall only be interested in properties of the $q$-Gaussian dependent on the varying parameter $q$, the parameter $\beta$ can be chosen arbitrarily; from now on we will consider $\beta = 1/(3-q)$. There are several reasons for this choice. First, when considering the $q$-generalization of variance of a $q$-Gaussian PDF:
\begin{equation}
\sigma_q^2 :=\int_{-\infty}^{\infty} x^2\frac{[f(x)]^q}{\int_{-\infty}^\infty [f(y)]^q dy} dx,
\label{eq::qgaussian.qvariance}
\end{equation}
one obtains a particularly simple result for the selected $\beta$, which is $\sigma_q^2=1$~\cite{UmarovS-2008a,deSantaHelenaEL-2015a}. Second, it is advantageous to consider $\sigma_q^2$ instead of the standard $\sigma^2 \equiv \sigma_{q=1}^2$ as it is finite and well defined for all $q \in (-\infty,3)$ unlike $\sigma^2$. Thus, the preferred choice of $\beta$ provides one with a standardized family of $q$-Gaussian distributions in the same way that taking $\sigma^2=1$ gives the standard normal distribution. Third, a more aesthetic reason is that this value of $\beta$ gives one a limit of the support of $\mathcal{G}_q$ as $q \to -\infty$, which is independent of $q$:
\begin{equation}
\lim_{q\to\-\infty} \mathbf{supp}(p_q) = [-1,1].
\end{equation}
In the above limit, the $q$-Gaussian PDF becomes uniform on $[-1,1]$: $p_{-\infty}(x)=1/2$ and it is zero outside this interval. This follows from the fact that for $x\in[-1,1]$ we have (see Appendix A):
\begin{equation}
\lim_{q\to-\infty} \frac{1}{\sqrt{3-q}C_q}\left(1-\frac{1-q}{3-q}x^2\right)^{\frac{1}{1-q}} = \frac{1}{2}.
\label{eq::qgaussian.minus-infinity}
\end{equation}
From the current perspective, it is important to note that, although the $q$-Gaussians are symmetric with respect to $x=0$, their key properties described above hold for unsigned random variables $x>0$ as well. The respective unsigned $q$-Gaussians will be denoted by $\mathcal{G}_q^{+}$ henceforth.


\section{Results}
\label{sect::results}

As it has already been mentioned in Sect.~\ref{sect::introduction}, multifractality in time series is often attributed in literature to two factors: temporal correlations and heavy tails of PDF as if these factors acted independently and each of them could contribute to an observed multifractal property of a data set under study~\cite{IvanovPC-1999a,KantelhardtJ-2002a,MatiaK-2003a,KwapienJ-2005b,BarunikJ-2012a,GomesLF-2023a}. We shall show that genuine multifractality stems \textit{only} from temporal correlations. While heavy tails can add to the complexity of the signal if the correlations are present, they cannot produce any multifractality on their own without said correlations.

In order to investigate what is a source of time series multifractal complexity, it is instructive to prepare two modified time series for a given original signal. The first time series is characterized by the same PDF as the source signal's one, but it lacks temporal correlations, which have been removed through random shuffling. This time series allows one to take into account only the effects of PDF heavy tails. The second time series has a modified PDF that differs substantially from the original signal's one, but its temporal organization has been preserved. This time series serves as a subject of analysis of how the correlations alone affect the fractal properties of the signal. This time series is used to determine the impact of correlation on the fractal properties of the original signal. The first of these two approaches was a subject of an earlier study~\cite{DrozdzS-2009a}, therefore this approach will be only briefly sketched here.

\subsection{Case 1: Heavy tails, no temporal structure of data}
\label{sect::heavy-tails}

We consider uncorrelated, unsigned time series with different degrees of tail heaviness parametrized by the parameter $q$ of the $q$-Gaussian PDFs: $1 \leqslant q \leqslant 2$. The extreme values of this interval correspond to the Gaussian PDF ($q=1$) and, in terms of the limiting distribution, to the Cauchy-Lorentz PDF ($q=2$). So-defined interval comprises both the Gaussian basin of attraction for $q < 5/3$ and the L\'evy-Gnedenko basin of attraction for $q \geqslant 5/3$. Of course, the q-Gaussian distributions do not align with the corresponding limiting distributions; however, their PDFs exhibit asymptotically power-law tails like in the case of the L\'evy-stable distributions (see Eq.~(\ref{eq::levy.qgaussian})).


\begin{figure}
\centering
\includegraphics[width=0.95\textwidth]{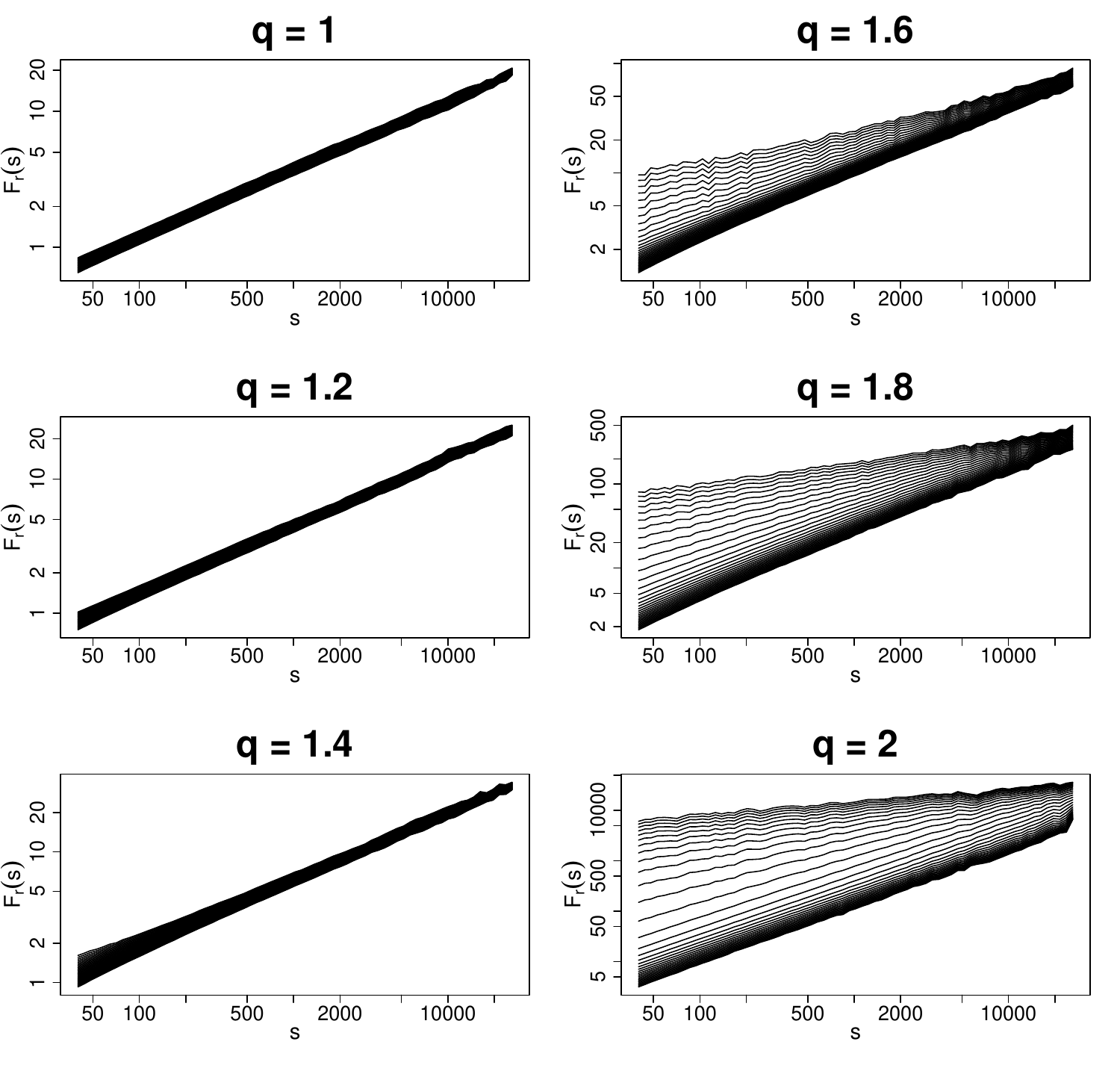}
\caption{Fluctuation functions $F_r(s)$ with $-4 \leqslant r \leqslant 4$ obtained by using MFDFA for time series of uncorrelated $q$-Gaussian noise with sample values of $q$. All the time series have a length of $N=10^5$ data points and the results have been averaged over 10 independent realizations of a corresponding process for a better clarity.}
\label{fig::qgaussian.uncorrelated.fluctuation-functions}
\end{figure}

In theory, uncorrelated random signals are monofractal if their distribution belongs to the Gaussian basin and bifractal if their distribution belongs to the L\'evy-Gnedenko basin~\cite{NakaoH-2000a,DrozdzS-2009a}. However, in real situations when finite-size effects are present, one cannot expect to obtain such a clear picture for uncorrelated signals. A reliability condition for MFDFA requires a good statistics both in the number of segments $M_s$ a studied signal is divided into (Eq.~(\ref{eq::fluctuation.functions})) and in the size $s$ of individual segments $\nu$, for which variance $f^2(\nu,s)$ is calculated (Eq.~(\ref{eq::variance})). Both requirements can be satisfied simultaneously if a time series is long enough so that $T/s \gg 1$ for $s \gg 1$. However, the more fat-tailed is the PDF of the time series, the harder these requirements are to be met, because the less reliable is calculation of variance for a given $s$. For $q > 5/3$, which corresponds to $\alpha_{\rm L} < 2$, this is even impossible in principle, because the related PDF do not have the second moment. This is why interpreting the results of a multifractal analysis in this case is a delicate issue requiring a lot of caution.

As calculation of variance is an inherent step of MFDFA, the method may nevertheless be applied even for $q > 5/3$ (see Ref.~\cite{DrozdzS-2009a}) but its results should then be interpreted more qualitatively than quantitatively. This refers especially to a situation, in which an obtained spectrum $f(\alpha)$ is broad but largely left-side asymmetric. For a time series with a heavy-tailed PDF, such a spectrum cannot be interpreted as multifractal, because, actually, it is a bifractal spectrum consisting of two points $(0,0)$ and $(\alpha_{\rm L},1)$, which is characteristic for L\'evy flights~\cite{NakaoH-2000a}, after being artificially broadened due to finite-size effects. This is a numerical artifact whose nature is the same as for a short monofractal time series, for which $f(\alpha)$ becomes a narrow parabola rather than the expected single point.


\begin{figure}
\centering
\includegraphics[width=0.9\textwidth]{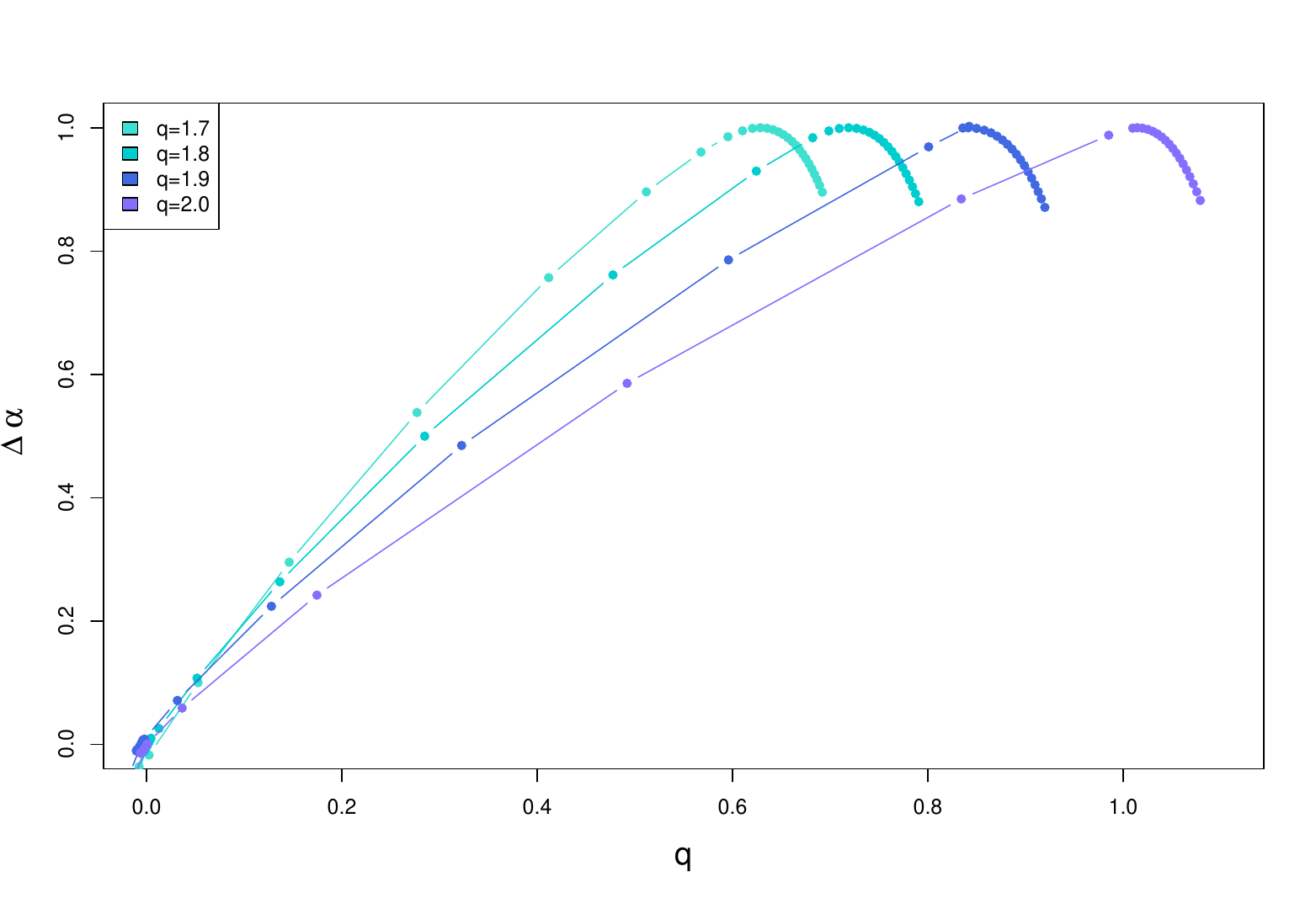}
\caption{Singularity spectra $f(\alpha)$ for time series of uncorrelated $q$-Gaussian noise with sample values of $q$ from the L\'evy-Gnedenko basin of attraction: $5/3 < q \leqslant 2$. The spectra have been calculated by using MFDFA with $-4 \leqslant r \leqslant 4$ and averaged over 10 independent realizations of a corresponding process.}
\label{fig::qgaussian.uncorrelated.falpha}
\end{figure} 

Another issue that can lead to misinterpretation of results from MFDFA has its origin in a slow convergence rate of non-Gaussian distributions with fat tails, which belong to the Gaussian basin. The fluctuation functions for sample values of $q$ are shown in Fig.~\ref{fig::qgaussian.uncorrelated.fluctuation-functions}. While for $q=1$ and $q=1.2$ one should see no ambiguity in reading and deciphering of the results, which clearly indicate a monofractal structure of the data, the picture is more complicated for $q=1.4$ and $q=1.6$, where one can identify two different regimes depending on a selected range of scales $s$. This is the case, in which a spurious identification of multiscaling can occur for small $s$. The observed broadening of the spectrum of $F_r(s)$ for the considered range of $q$, which becomes more significant as $q$ increases, actually arises from a slow convergence of the PDF to its Gaussian attractor. One only needs to look at sufficiently large scales in order to avoid this unfavorable effect. This problem is no longer faced for $q=1.8$ and $q=2$, which correspond to the L\'evy-Gnedenko basin of attraction. Here the spread-out fluctuation functions maintain their behavior across the entire range of scales with no cross-over point observed. In this region of $q > 5/3$, the singularity spectra $f(\alpha)$ derived with MFDFA are bifractal, but in practice these look like in Fig.~\ref{fig::qgaussian.uncorrelated.falpha}. Note the shift of the maximum of $f(\alpha)$ as $q$ increases, in agreement with increasing $\alpha_{\rm L}$ described by Eq.~(\ref{eq::levy.qgaussian}). A relative scarcity of the $f(\alpha)$ values between the accumulation regions near $\alpha=0$ and $\alpha=\alpha_{\rm L}$ is an indication of a bifractal nature of the results.


\begin{figure}
\centering
\includegraphics[width=0.9\textwidth]{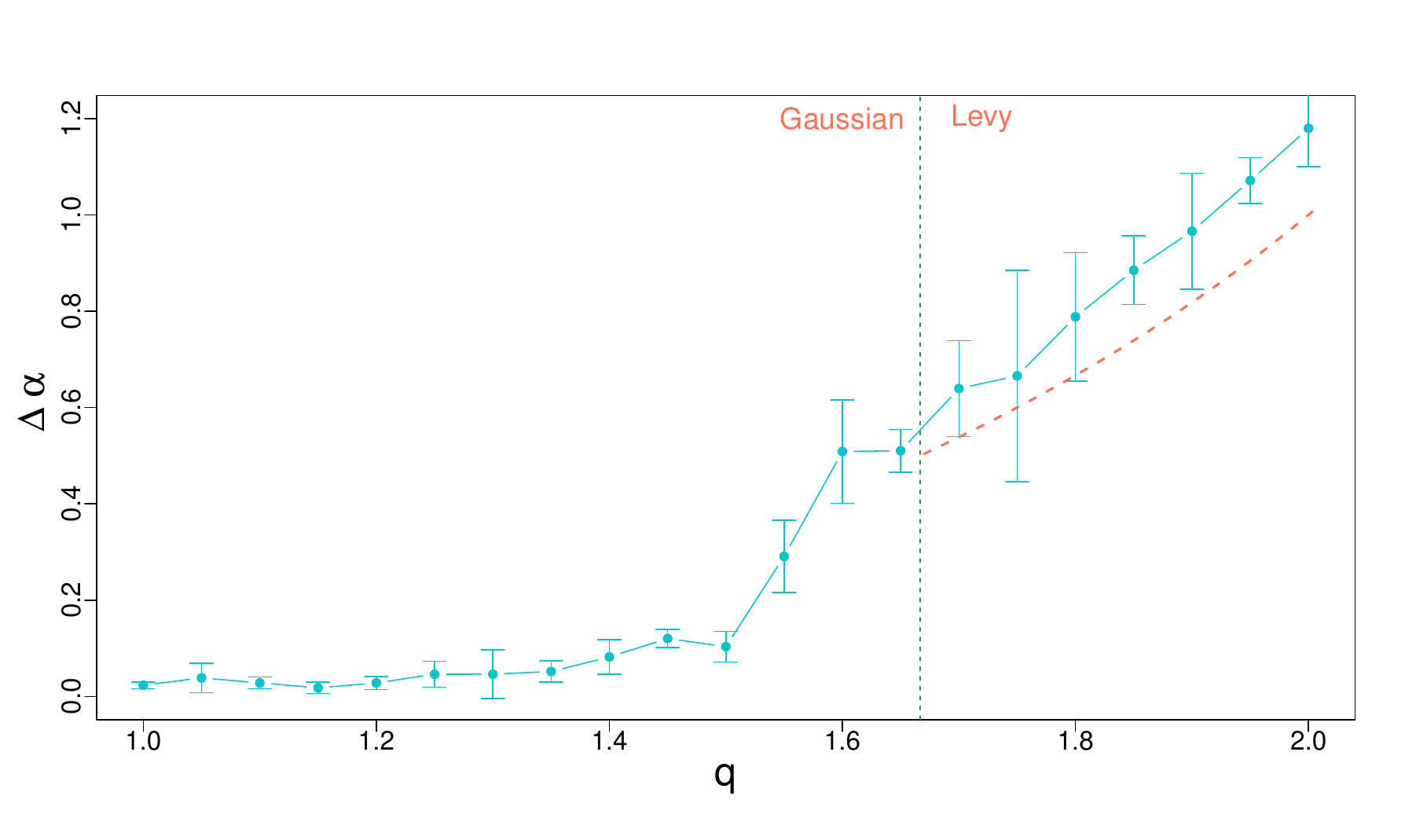}
\caption{Width $\Delta\alpha$ of the singularity spectra $f(\alpha)$ calculated for time series of uncorrelated $q$-Gaussian noise as a function of $q$ in the range $1 \leqslant q \leqslant 2$. Vertical line at $q=5/3$ separates the Gaussian and L\'evy-Gnedenko basins. In the latter basin, the orange dashed line denotes the theoretical spectrum width $\Delta\alpha = 1/\alpha_{\rm L}$ for bifractal signals with the same $q$. All the time series have a length of $N=10^5$ data points and the results have been averaged over 10 independent realizations of a corresponding process. Error bars denote standard deviation of $\Delta\alpha$.}
\label{fig::qgaussian.falpha.width}
\end{figure} 

The problems with properly estimating the singularity spectra $f(\alpha)$ that have been discussed above can be presented in a quantitative form by plotting a dependence of the width of $f(\alpha)$ on the Tsallis parameter $q$ for a range of its values. Fig.~\ref{fig::qgaussian.falpha.width} shows this dependence for $1 \leqslant q \leqslant 2$, which comprises both the Gaussian basin for $q < 5/3$ and the L\'evy-Gnedenko basin for $q \geqslant 5/3$. Despite the fact that, in the latter case, calculation of variance is not recommended from a theoretical perspective, the results show that MFDFA allows one for a relatively reasonable estimation of $\Delta\alpha$, which is only little larger than its analytical value $\Delta\alpha=1/\alpha_{\rm L}$~\cite{NakaoH-2000a}. For $q < 1.5$ the results are in agreement with the expected monofractal structure of the corresponding processes ($\Delta\alpha \approx 0$). The only problematic range of $q$ is between 1.5 and 5/3, where the numerical procedure ``senses'' the heavy tails but the length of the time series is too small to obtain a clear convergence predicted by the central limit theorem. The width of this range can be narrowed by extending the time series, because one expects that $\Delta\alpha \to 0$ with $T \to \infty$ (see Ref.~\cite{DrozdzS-2009a} for a related discussion).

\subsection{Case 2: Preserved temporal correlations, distorted PDF tails}
\label{sect::temporal-correlations}

How a co-occurrence of the fat-tail PDFs and the temporal correlations impacts the multifractal properties of time series one can infer from an analysis, in which PDFs of the time series under study have been distorted in a controlled manner, while the temporal organization of the data points has been preserved. It must be stressed, however, that preserving the temporal organization of the data does not mean that the correlation structure of the modified time series is exactly the same as the structure of the original time series: by reshaping a PDF, one distorts the correlations, but not remove them.

The PDF of a source time series $U$ can be replaced by a target PDF by using a ranking-based probability density transformation. In this study, the $q$-Gaussian distributions $\mathcal{G}_q^{+}$ with varying $q$ are selected for the target PDFs. The PDF-reshaping transformation is started with rank-ordering of the original time series $U=\{u_i\}_{i=1}^T$ by a mapping $\mathcal{R}: U \to R = \{R_i\}_{i=1}^T$, where $R_i = R_1+\#\{u_j \in U: u_j > u_i\}$ and $\#$ denotes cardinality of the set (if $m$ data points are attributed with the same rank $R_k$, their ranks are changed randomly to $R_{k'}=R_{k+l}$ with $l=0,...,m-1$). In parallel, a new time series $G_q=\{g_i\}_{i=1}^T$, with $g_i > 0$ for all $i$, is generated from a given $q$-Gaussian PDF and also rank-ordered by $\mathcal{R'}: G_q \to R' = \{R_i^{'}\}_{i=1}^T$. Then, $R'$ is transformed into a target time series by $\mathcal{R}^{-1}: R' \to \overset{\sim}{G_q} = \{\tilde{g}_i\}_{i=1}^T$ so that the entire transformation can be written as
\begin{equation}
\overset{\sim}{G}_q=\mathcal{R}^{-1}(\mathcal{R}'(G_q)).
\label{eq::rank-ordering.transformation}
\end{equation}
The time series $\overset{\sim}{G}_q$ has a $q$-Gaussian distribution with a given $q$, but its temporal organization is inherited from the source time series $U$. In the following, $\overset{\sim}{G}_q$ will be subject to multifractal analysis for different values of $q$. Starting from a Gaussian-distributed time series ($q=1$), by gradually increasing or decreasing mass of the PDF tails, one may then investigate the impact of PDF tails on the time-correlated data. Time series representing multiplicative cascades of different types discussed in Sect.~\ref{sect::multiplicative.cascades} will be used as the source signals. A broad range of the Tsallis parameter will be considered: $q\in(-\infty,2]$.

\subsubsection{$q$-Gaussian with a binomial-cascade organization}

\begin{figure}
\centering
\subfloat[]{\includegraphics[trim={0 20 0 0},clip,width=0.85\textwidth]{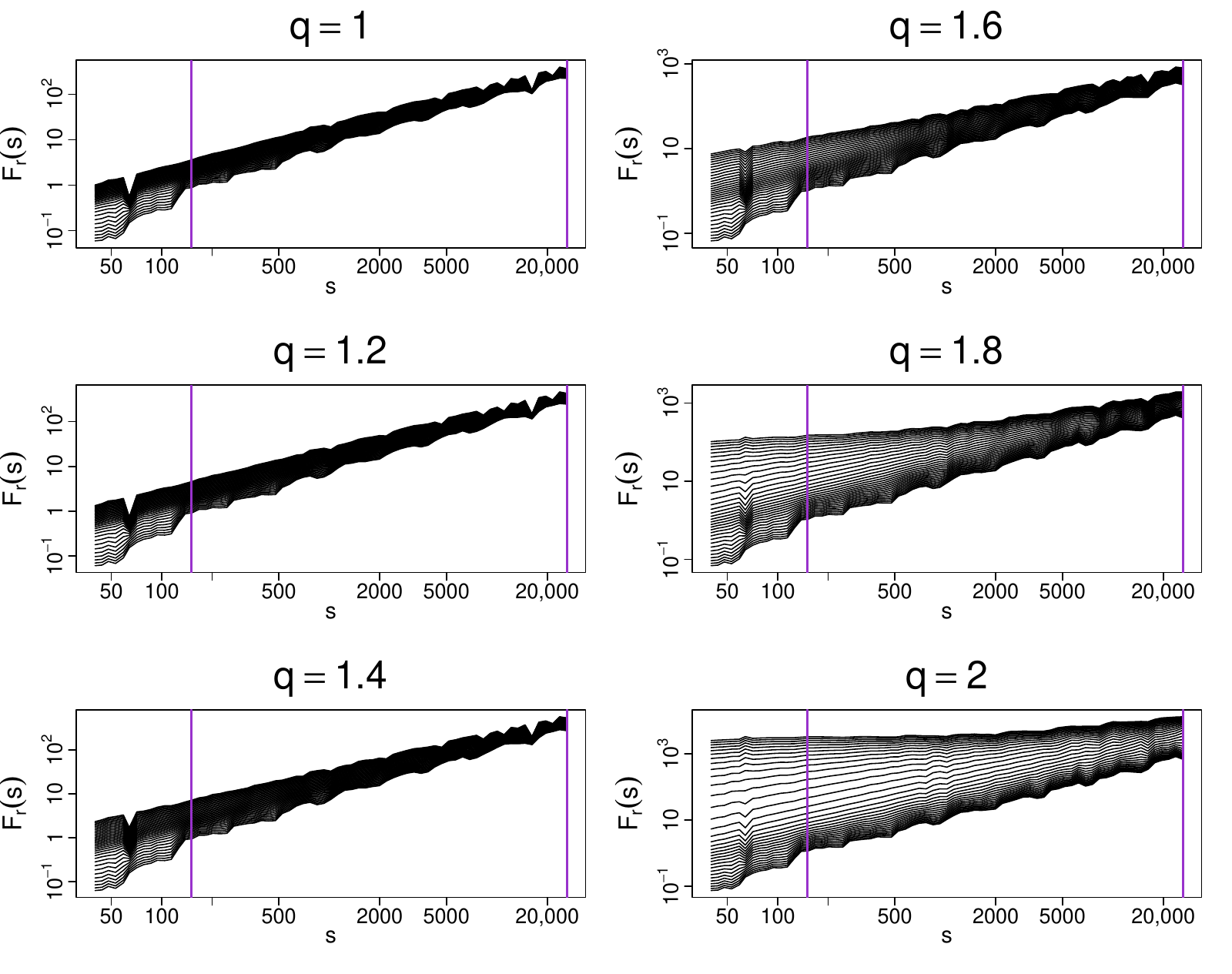}}

\vspace{0.2cm}
\subfloat[]{\includegraphics[trim={0 20 0 0},clip,width=0.85\textwidth]{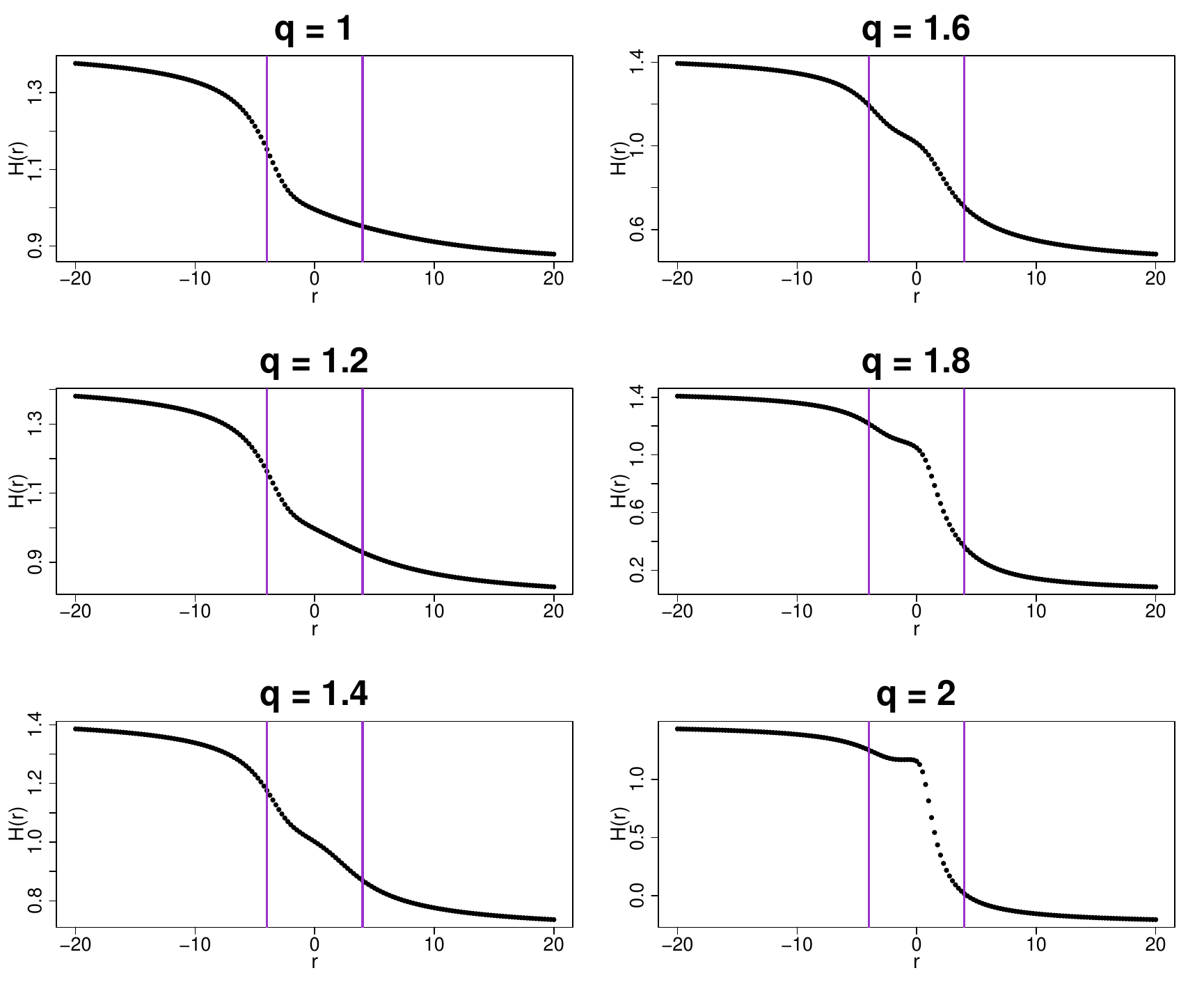}}
\caption{(a) Fluctuation functions $F_r(s)$ and (b) the generalized Hurst exponents $h(r)$ calculated by using MFDFA from $q$-Gaussian time series with $1 \leqslant q \leqslant 2$ and their temporal organization inherited from a dyadic binomial cascade. Vertical lines denote the lower and upper boundaries of (a) the range of scales $s$ used to calculate $h(r)$ and (b) the range $-4 \leqslant r \leqslant 4$ used to calculate singularity spectra $f(\alpha)$ shown in Fig.~\ref{fig::transformed.binomial.heavy-tails.falpha}. For $q=1$ the time series has a Gaussian PDF, for $q=1.2$, $q=1.4$, and $q=1.6$, the PDFs exhibit heavy tails but their second moment is finite, while for $q=1.8$ and $q=2$, the tails are so heavy that the second moment does not exist. The results have been averaged over 10 independent realizations of the $q$-Gaussian time series.}
\label{fig::transformed.binomial.heavy-tails.Frh}
\end{figure}

\begin{figure}
\centering
\includegraphics[width=0.9\textwidth]{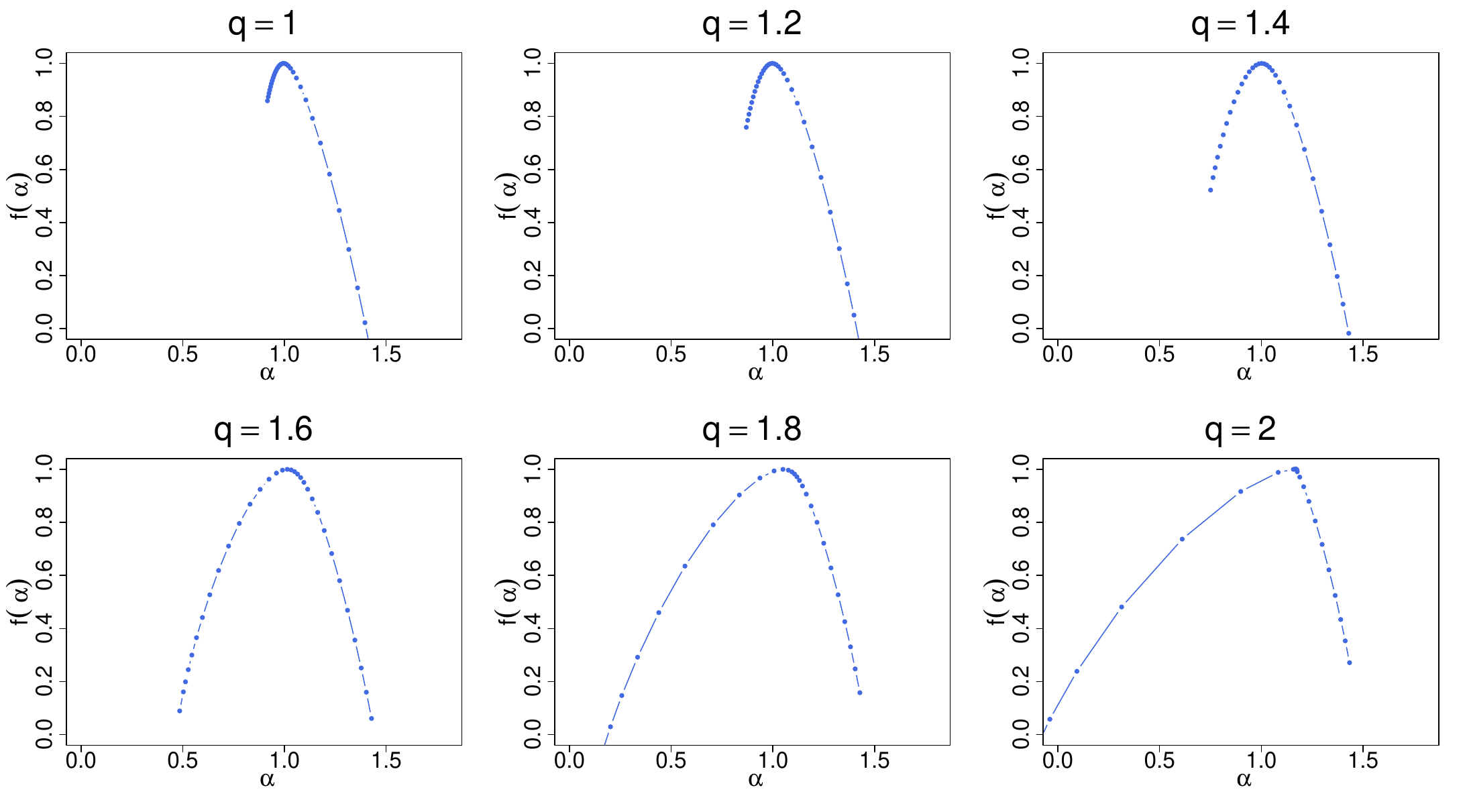}
\caption{Singularity spectra $f(\alpha)$ for the same time series as in Fig.~\ref{fig::transformed.binomial.heavy-tails.Frh}. The results have been averaged over 100 independent realizations of the $q$-Gaussian time series.}
\label{fig::transformed.binomial.heavy-tails.falpha}
\end{figure}

Let the analysis be started with a deterministic binomial cascade with $k$ iterations and a multiplier that meets the condition $0.5 < p < 1$. The exact value of $p$ is unimportant from the present perspective, because rank-ordering of such a time series does not alter if $p$ is varied. The procedure described above is applied to this cascade, which produces a family of the transformed time series $\overset{\sim}{G_q^{\rm B}}$ with $q$-Gaussian PDFs. Fig.~\ref{fig::transformed.binomial.heavy-tails.Frh}(a) shows the fluctuation functions $F_r(s)$ calculated from these time series for a few sample values of $q$ in the range $1 \leqslant q \leqslant 2$. What is the most interesting here, is that the temporal organization of the binomial cascade for $q=1$ produces a clear spread of the fluctuation functions characteristic for the multifractals even though the analyzed time series have Gaussian PDFs. This can be verified in Fig.~\ref{fig::transformed.binomial.heavy-tails.falpha}, where the respective spectrum is broad ($\Delta\alpha>0.4$) and strongly right-side asymmetric. In this case, multifractality comes from small fluctuations as the right shoulder of $f(\alpha)$ represents $r<0$.


By increasing $q$, the functions $F_r(s)$ tend to spread out more and more, which is reflected in the width of their $f(\alpha)$. The right-side asymmetry gradually disappears and for $q>1.5$ the left shoulder starts to dominate (Fig.~\ref{fig::transformed.binomial.heavy-tails.falpha}). It is recommended to juxtapose the fluctuation functions shown in Figs.~\ref{fig::qgaussian.uncorrelated.fluctuation-functions} and~\ref{fig::transformed.binomial.heavy-tails.Frh}. What differentiates between these two cases, is the lack of temporal correlations in the former and their presence in the latter. As for the standard Gaussian PDF, the introduction of mixed linear and nonlinear correlations changes the fractal structure from a monofractal one to a multifractal one. If the larger values of $q$ are considered, $1 < q < 5/3$, the correlations shift ``the small-scale broom effect'' from $F_r(s)$ with $r>0$ (i.e., from large fluctuations) to $F_r(s)$ with $r<0$ (i.e., to small fluctuations) and produce a homogeneous scaling of $F_r(s)$ over a substantially wide range of $s$. Such a homogeneity of the scaling may be interpreted as a signature of genuine multifractality~\cite{DrozdzS-2009a}. The presence of temporal correlations has also an impact on the results for $q > 5/3$. Now the distributions belong to the L\'evy-Gnedenko basin and the analyzed time series resemble L\'evy flights with long-range memory. The correlations distort the bifractal structure of the memoryless L\'evy flights so indisputably manifesting itself in the $f(\alpha)$ spectra in Fig.~\ref{fig::qgaussian.uncorrelated.falpha} and produce their broadening. For $q=1.8$ and $q=2$ in Fig.~\ref{fig::transformed.binomial.heavy-tails.Frh}(a), $f(\alpha)$ no longer consists of two points, but nevertheless the spectra still ``remember'' the model bifractal shape with the right-side asymmetry. As an addition, the generalized Hurst exponent functions $h(r)$ are displayed in Fig.~\ref{fig::transformed.binomial.heavy-tails.Frh}(b). Their characteristic feature is an unconventional shape observed near $r=0$, where a concave bump is located, whose prominence grows with $q$. Its existence doesn't alter the provided interpretation of the results, though.


\begin{figure}
\centering
\includegraphics[width=0.9\textwidth]{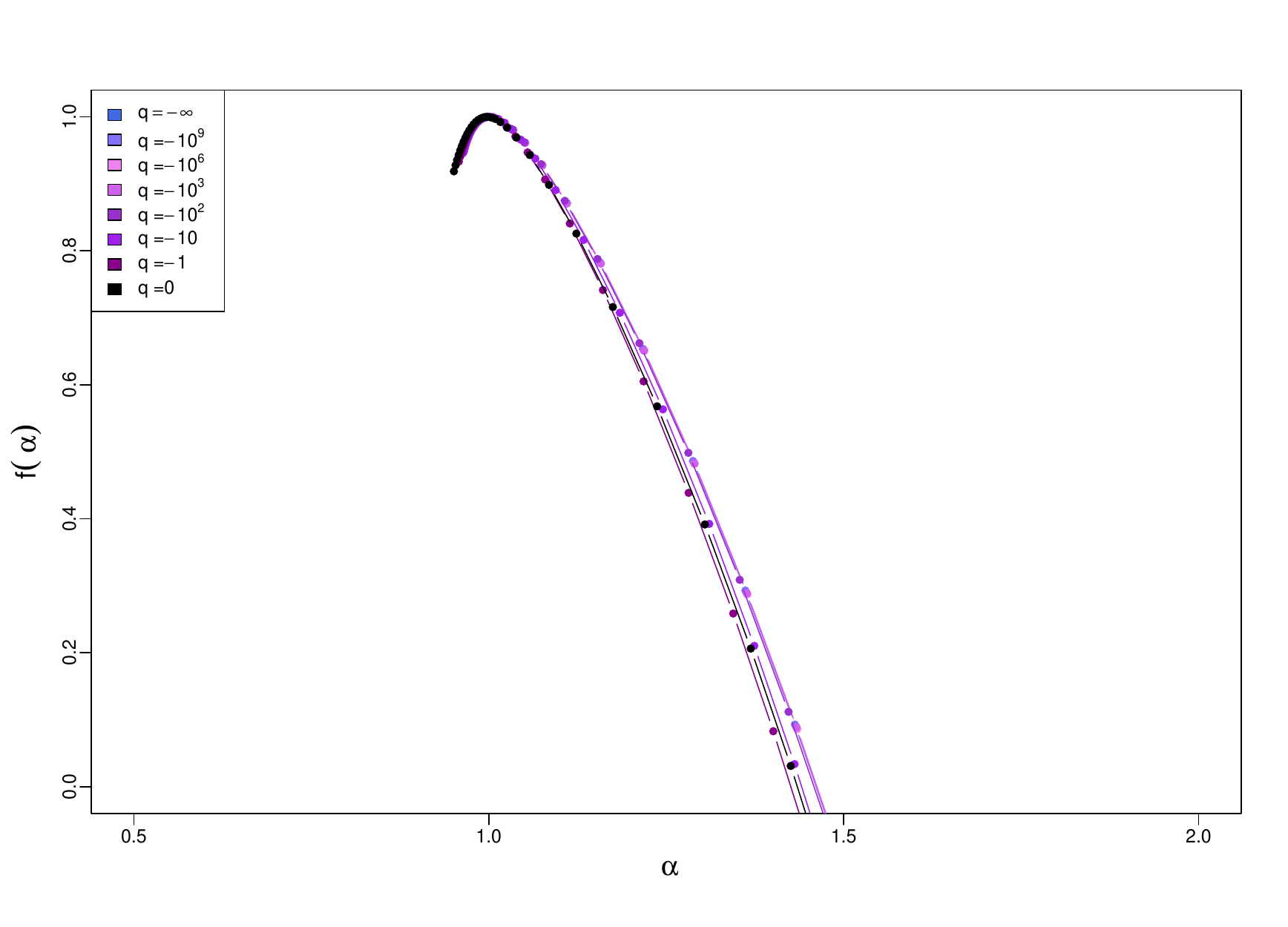}
\caption{Singularity spectra $f(\alpha)$ for time series with $q$-Gaussian PDFs on a compact support, where $-\infty < q < 1$, and with temporal organization inherited from a dyadic binomial cascade. The results have been averaged over 10 independent realizations of the $q$-Gaussian time series.}
\label{fig::transformed.binomial.no-tail.falpha}
\end{figure}

Fig.~\ref{fig::transformed.binomial.no-tail.falpha} collects the singularity spectra calculated for the time series whose PDFs have a compact support, which correspond to $q < 1$. Obviously, all those PDFs belong to the Gaussian basin of attraction. It occurs that their particular shape does not impact the results. Neither the functions $F_r(s)$ nor $h(r)$ differ from their counterparts for the Gaussian case of $q=1$. Therefore, its is sufficient to show only the respective $f(\alpha)$ here in order to illustrate this statement.

\begin{figure}
\centering
\includegraphics[width=0.9\textwidth]{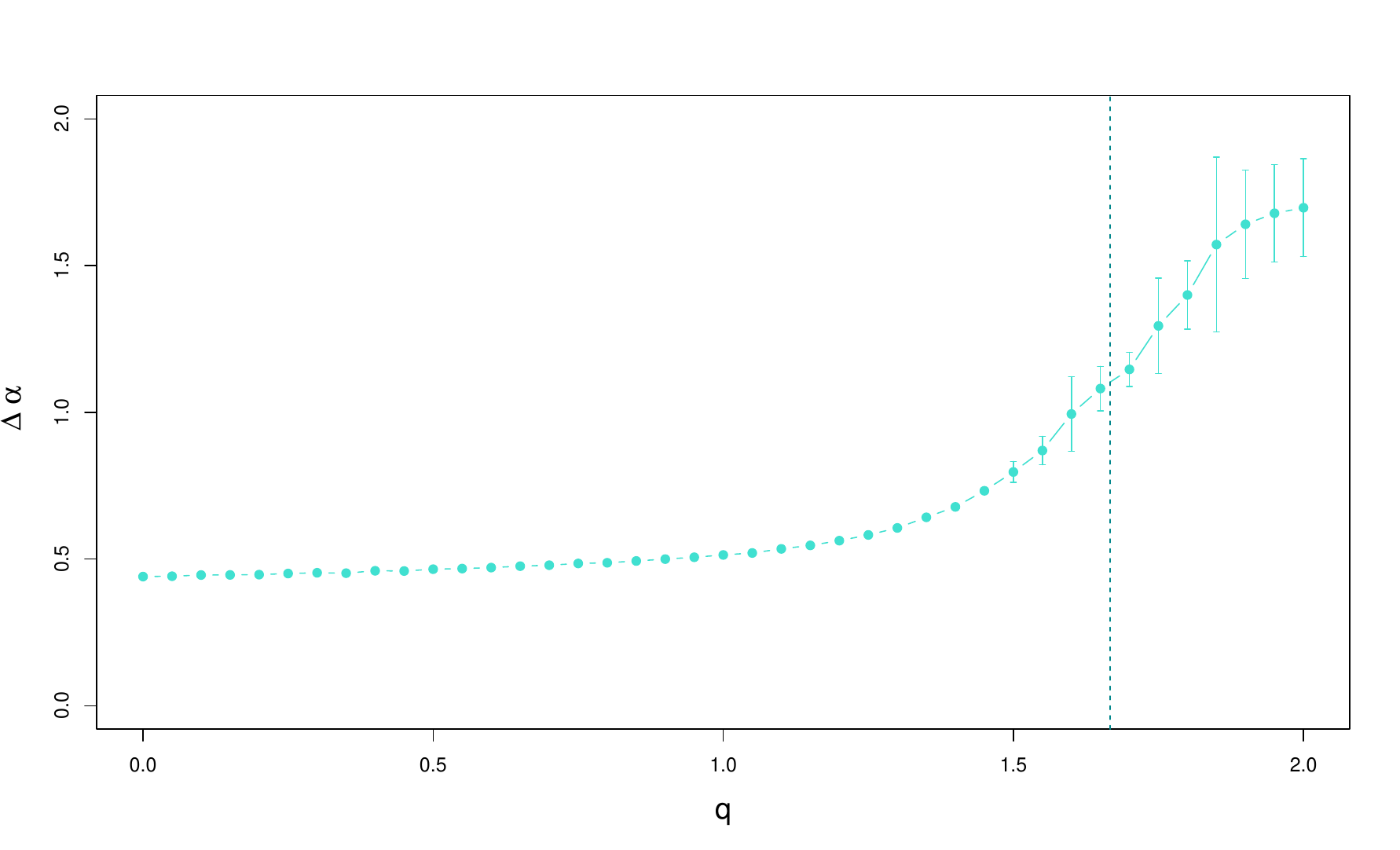}
\caption{Width $\Delta\alpha$ of the singularity spectra $f(\alpha)$ calculated for $q$-Gaussian time series of length of $N=10^5$ data points with $0 \leqslant q \leqslant 2$ and their temporal organization inherited from a dyadic binomial cascade. Vertical line at $q=5/3$ separates the Gaussian and L\'evy-Gnedenko basins. Error bars denote standard deviation of $\Delta\alpha$ calculated from 20 independent realizations of the $q$-Gaussian time series.}
\label{fig::transformed.binomial.falpha.width}
\end{figure}

In order to present the characteristics of $f(\alpha)$ for a broad range of $q$, a functional dependence of $\Delta\alpha$, averaged over a number of independent realizations of the $q$-Gaussian time series, is shown if Fig.~\ref{fig::transformed.binomial.falpha.width}. One has to notice here the existence of two regimes of the variability of $\Delta\alpha$. In the first one located essentially below $q \approx 1.2$ (which depends on time series length, see Ref.~\cite{DrozdzS-2009a}), which $-$ in the absence of temporal correlations $-$ would correspond to $q$ deep below the boundary between the Gaussian and L\'evy-Gnedenko basins of attraction, the singularity spectrum width $\Delta\alpha \approx 0.5$ is not only stable over $q$ with only minor upward drift, but also remarkably stable across samples (see the negligible size of error bars denoting standard deviation). This picture is no longer valid when $q$ is increased further and exceeds 1.2. The slope of the functional dependence of the spectrum width on $q$ starts to increase and, for a sizable range of $q$ up to 1.85, it is almost linear with increasing variability across samples. Finally, for $q \geqslant 1.9$ a trace of saturation of $\Delta\alpha$ can be seen. Such a behavior of this quantity suggests that the crucial role in stability of $f(\alpha)$ plays the central limit theorem even if the correlations are present and responsible for the observed multifractality (which is associated with $\Delta\alpha \gg 0$). If the PDF tails are thin or its support is compact, in the no-correlation case, one would deal with a fast convergence to the limiting Gaussian distribution and the shape of $f(\alpha)$ would be expected to be point-like. Here the correlations broaden the spectrum, but they nevertheless seem to be unable to vary $f(\alpha)$ when one considers different samples. However, when the tails start to become heavy, the picture changes. In the no-correlation case, the convergence imposed by the central limit theorem would be slower, even though still valid. The growing error bars that show the variability of $\Delta\alpha$ between samples indicate that, again, the PDF shape is the crucial factor that determines stability of $f(\alpha)$. For $q>5/3$, in the L\'evy-Gnedenko regime, the attractor becomes one of the L\'evy-stable distributions, so the width of $f(\alpha)$ must reflect this fact also when the temporal correlations are present -- see Fig.~\ref{fig::transformed.binomial.falpha.width}.


\subsubsection{$q$-Gaussian with a log-normal-cascade organization}

\begin{figure}
\centering
\subfloat[]{\includegraphics[trim={0 20 0 0},clip,width=0.85\textwidth]{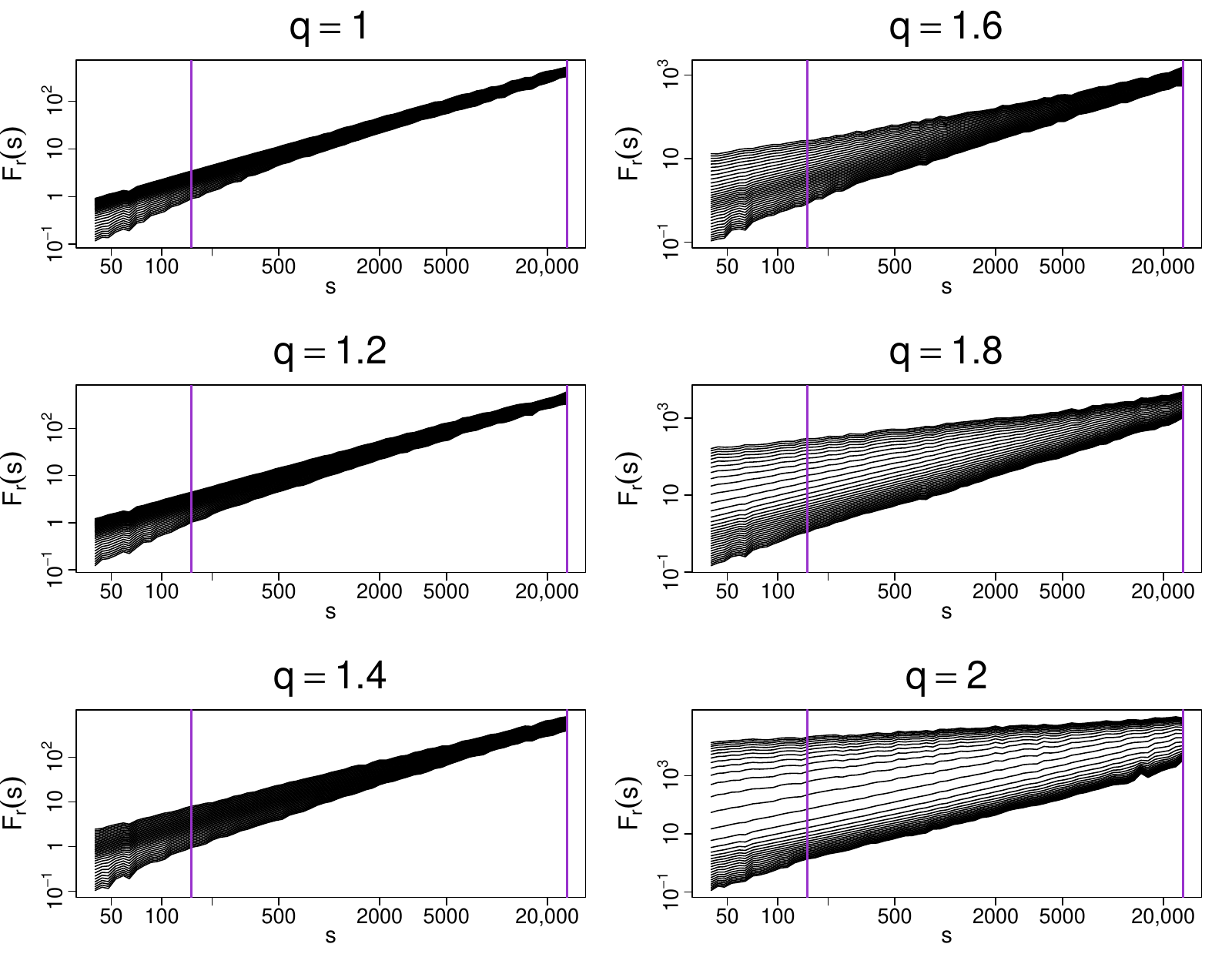}}

\vspace{0.2cm}
\subfloat[]{\includegraphics[trim={0 20 0 0},clip,width=0.85\textwidth]{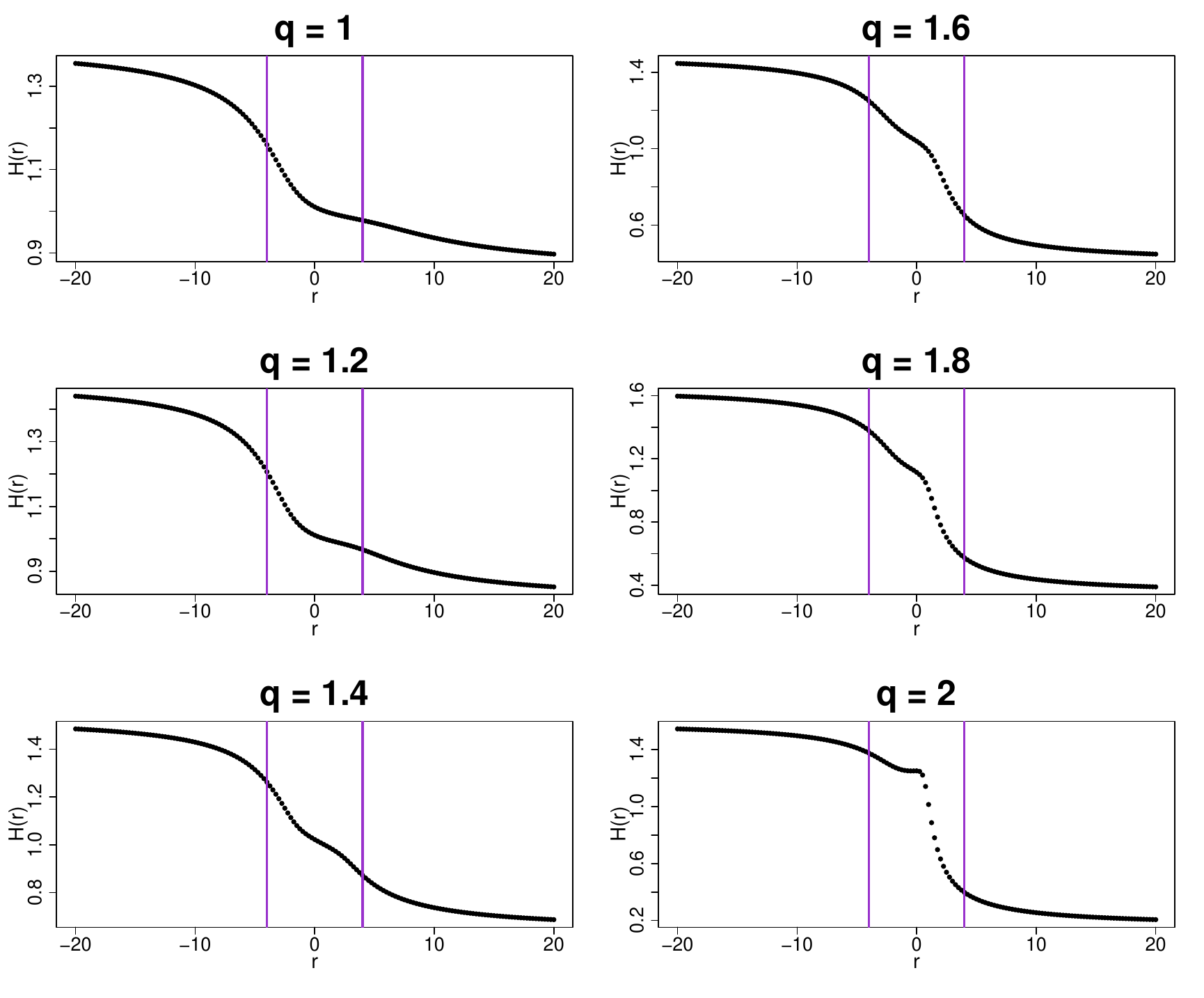}}
\caption{(a) Fluctuation functions $F_r(s)$ and (b) the generalized Hurst exponents $h(r)$ calculated by using MFDFA from $q$-Gaussian time series with $1 \leqslant q \leqslant 2$ and their temporal organization inherited from a dyadic log-normal cascade. Vertical lines denote the lower and upper boundaries of (a) the range of scales $s$ used to calculate $h(r)$ and (b) the range $-4 \leqslant r \leqslant 4$ used to calculate singularity spectra $f(\alpha)$ shown in Fig.~\ref{fig::transformed.log-normal.heavy-tails.falpha}. The results have been averaged over 10 independent realizations of the $q$-Gaussian time series.}
\label{fig::transformed.log-normal.heavy-tails.Frh}
\end{figure}

\begin{figure}
\centering
\includegraphics[width=1\textwidth]{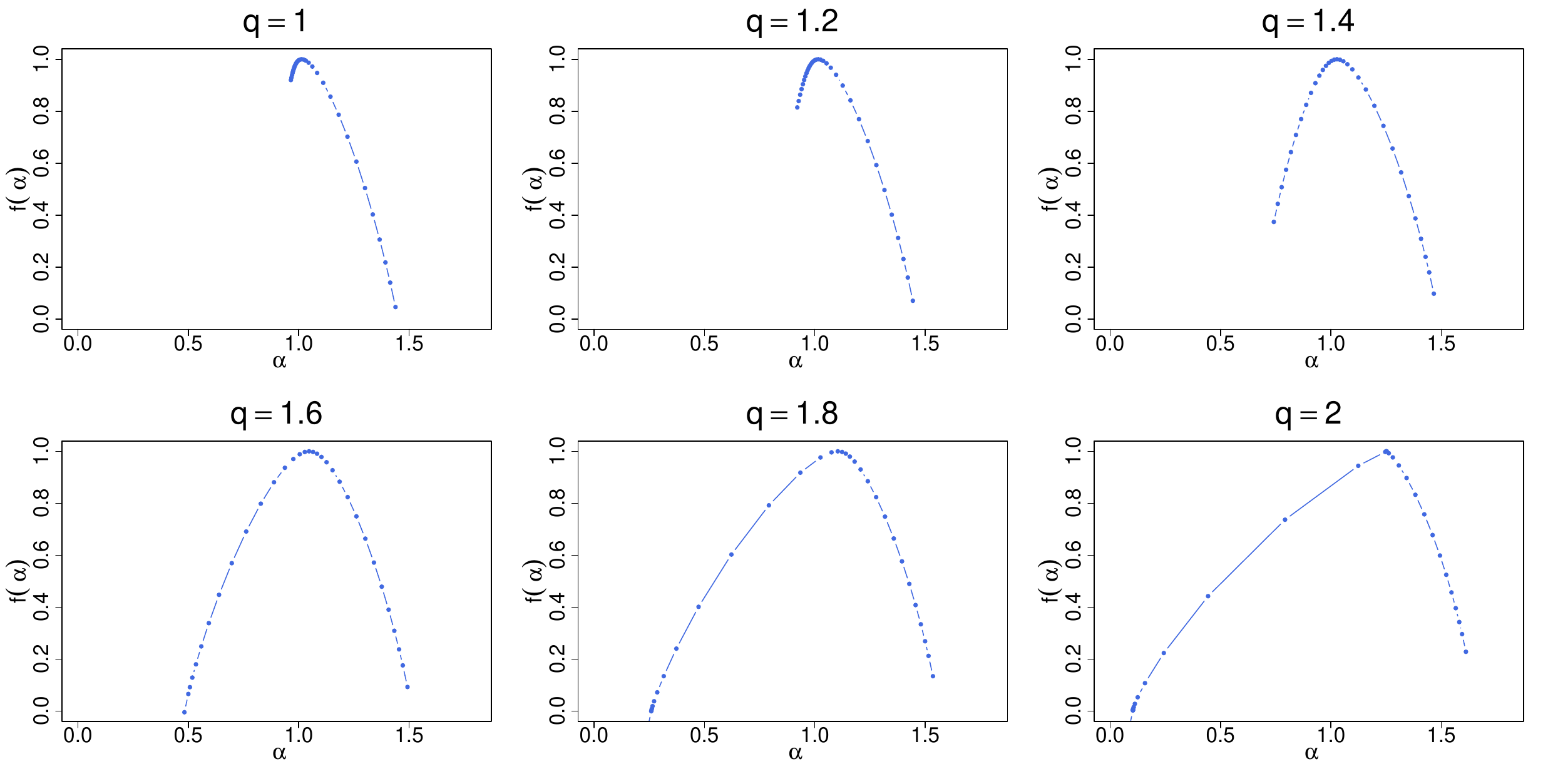}
\caption{Singularity spectra $f(\alpha)$ calculated from the same time series as the quantities shown in Fig.~\ref{fig::transformed.log-normal.heavy-tails.Frh}. The results have been averaged over 100 independent realizations of the $q$-Gaussian time series.}
\label{fig::transformed.log-normal.heavy-tails.falpha}
\end{figure}

The first canonical cascade to be considered here is the dyadic log-normal one with $k=17$ iterations. It forms a correlated time series, which is then transformed to a set of time series with $q$-Gaussian PDFs for different choices of $q$ by using the transformation from Eq.~(\ref{eq::rank-ordering.transformation}). If compared with the binomial-cascade case discussed above, now the fluctuation functions reveal a slightly worse scaling with more deflections from ideal straight lines $-$ see Fig.~\ref{fig::transformed.log-normal.heavy-tails.Frh}(a). Despite this fact, one still observes broadening of the corresponding $f(\alpha)$ spectra relatively to the no-correlation case presented in Fig.~\ref{fig::qgaussian.uncorrelated.fluctuation-functions}. When $q$ increases, this broadening becomes clearly larger with a preserved scaling over a wide range of scales $s$. This effect does not differ from the one observed for the binomial cascades and its interpretation regarding its origin must be the same. The generalized Hurst exponents are shown in Fig.~\ref{fig::transformed.log-normal.heavy-tails.Frh}(b) for an additional support of the conclusion that there is no qualitative difference between the outcomes for the deterministic and stochastic cascades. Indeed, the singularity spectra shown in Fig.~\ref{fig::transformed.log-normal.heavy-tails.falpha} closely resemble those shown in Fig.~\ref{fig::transformed.binomial.heavy-tails.falpha}.


\begin{figure}
\includegraphics[width=0.9\textwidth]{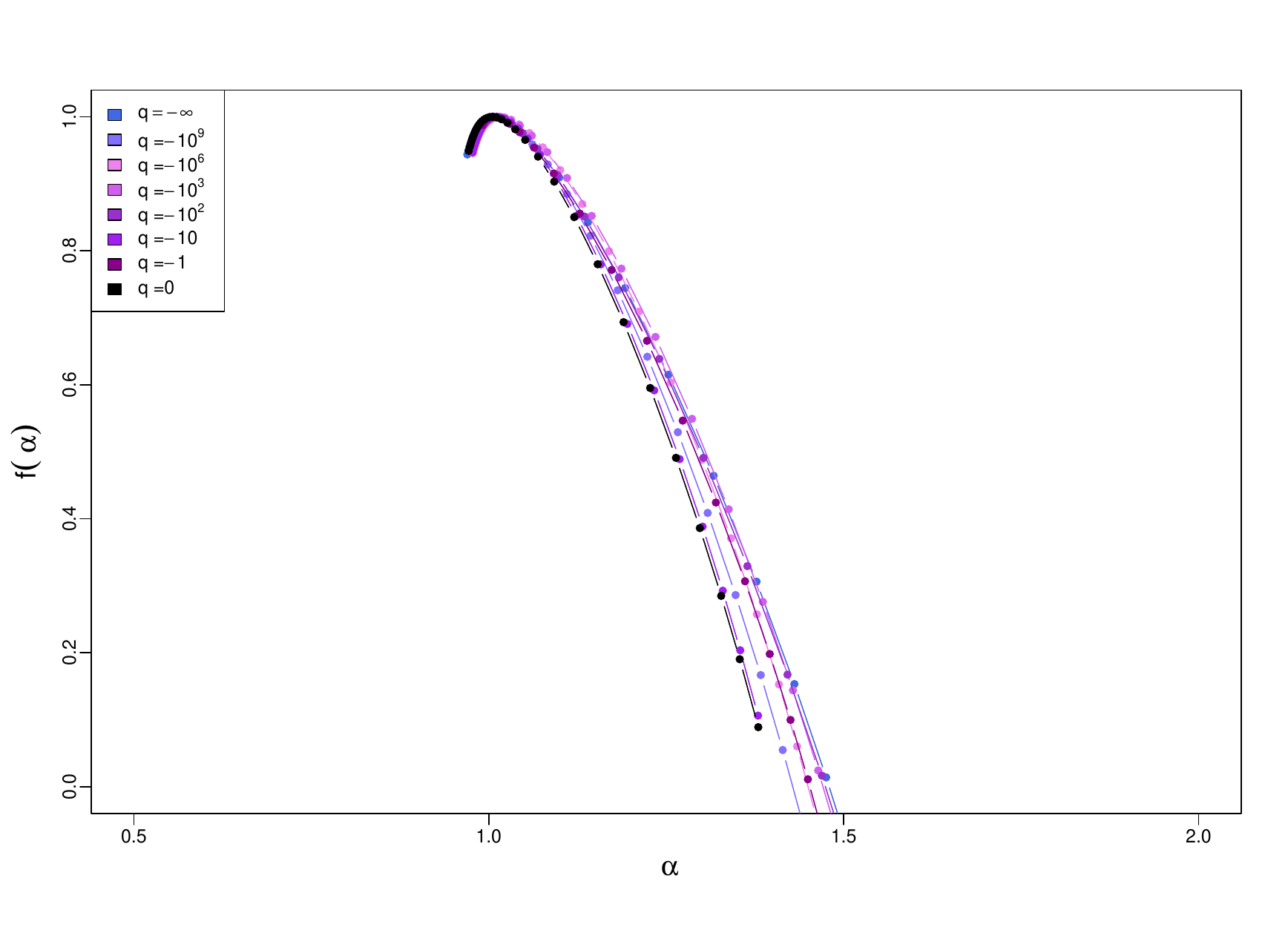}
\caption{Singularity spectra $f(\alpha)$ for time series with $q$-Gaussian PDFs on a compact support, where $-\infty < q < 1$, and with temporal organization inherited from a dyadic log-normal cascade. The results have been averaged over 10 independent realizations of the $q$-Gaussian time series.}
\label{fig::transformed.log-normal.no-tail.falpha}
\end{figure}

\begin{figure}
\includegraphics[width=0.9\textwidth]{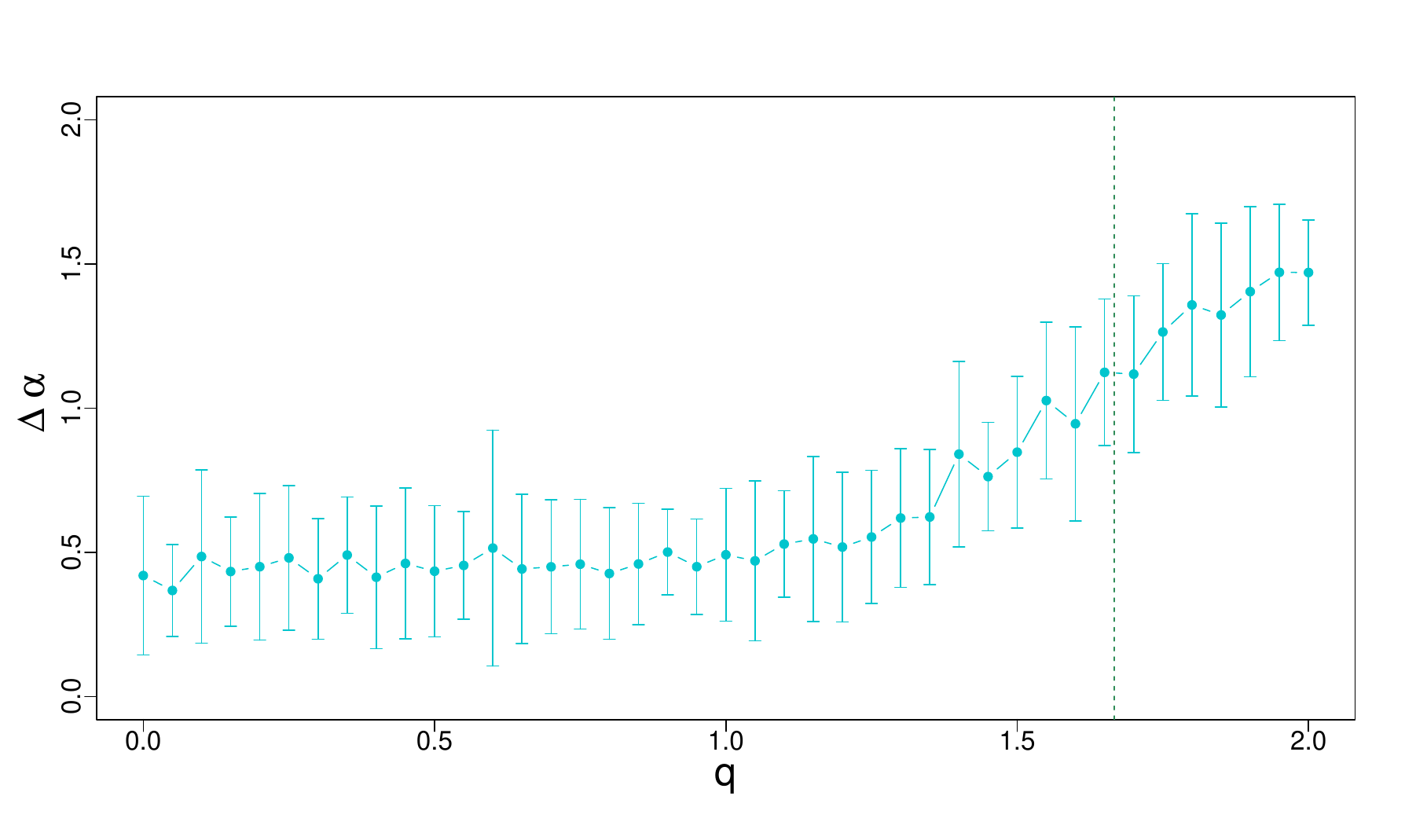}
\caption{Width $\Delta\alpha$ of the singularity spectra $f(\alpha)$ calculated for $q$-Gaussian time series of length of $N=10^5$ data points with $0 \leqslant q \leqslant 2$ and their temporal organization inherited from a dyadic log-normal cascade. Vertical line at $q=5/3$ separates the Gaussian and L\'evy-Gnedenko basins. Error bars denote standard deviation of $\Delta\alpha$ calculated from 20 independent realizations of the $q$-Gaussian time series.}
\label{fig::transformed.log-normal.falpha.width}
\end{figure}

A more sizable effect can be found for time series with the compact-support PDFs for $q<1$ as Fig.~\ref{fig::transformed.log-normal.no-tail.falpha} documents. The time series that inherited their temporal organization from the log-normal cascades show a grouping of the $f(\alpha)$ functions into two sets with slightly different widths $\Delta\alpha$. It is difficult to provide a decisive explanation for this effect. However, it is rather minor and cannot be considered as qualitative. Fig.~\ref{fig::transformed.log-normal.falpha.width} presents how the width of $f(\alpha)$ depends on the Tsallis parameter $q$ in a wide range of its variability. One can point out to two distinct regimes of its behavior. In the first one, which extends over the interval $0 \leqslant q \leqslant 1.2$, the quantity $\Delta\alpha$ is constant with its expectation value calculated from 20 independent realizations of the $q$-Gaussian processes oscillating between 0.4 and 0.5. This value must be interpreted as an indication of multifractality that, on average, is invariant over $q$ within the aforementioned range. Nonetheless, the variability of $\Delta\alpha$ between the individual realizations of the process with a given value of $q$ is substantial as its standard deviation denoted by error bars shows. This means that the multifractal complexity of the time series varies from sample to sample. One may attribute this outcome to a high variability of amplitude in the log-normal cascades (see Fig.~\ref{fig::binomial.cascade}).


Another regime of the dependence of $\Delta\alpha$ on $q$ is found for $q > 1.3$. In this regime, the spectrum width strongly increases with $q$ from 0.5 up to 1.5, which means that the multifractal complexity of the respective time series increases on average as well. The strong variability of $\Delta\alpha$ among the samples is also observed in this case.


\subsubsection{$q$-Gaussian with a log-gamma-cascade organization}

\begin{figure}
\centering
\subfloat[]{\includegraphics[trim={0 20 0 0},clip,width=0.9\textwidth]{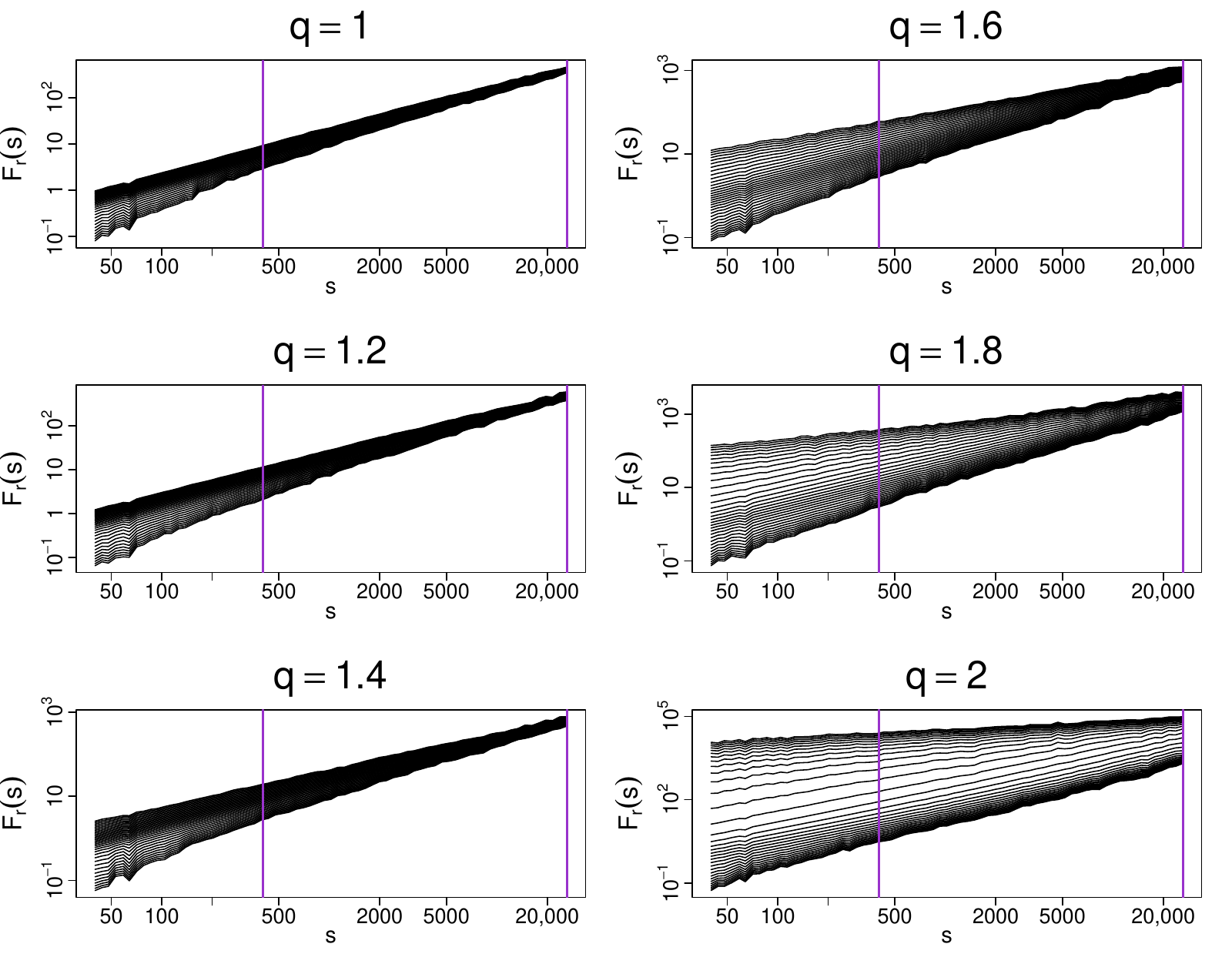}}

\subfloat[]{\includegraphics[trim={0 20 0 0},clip,width=0.9\textwidth]{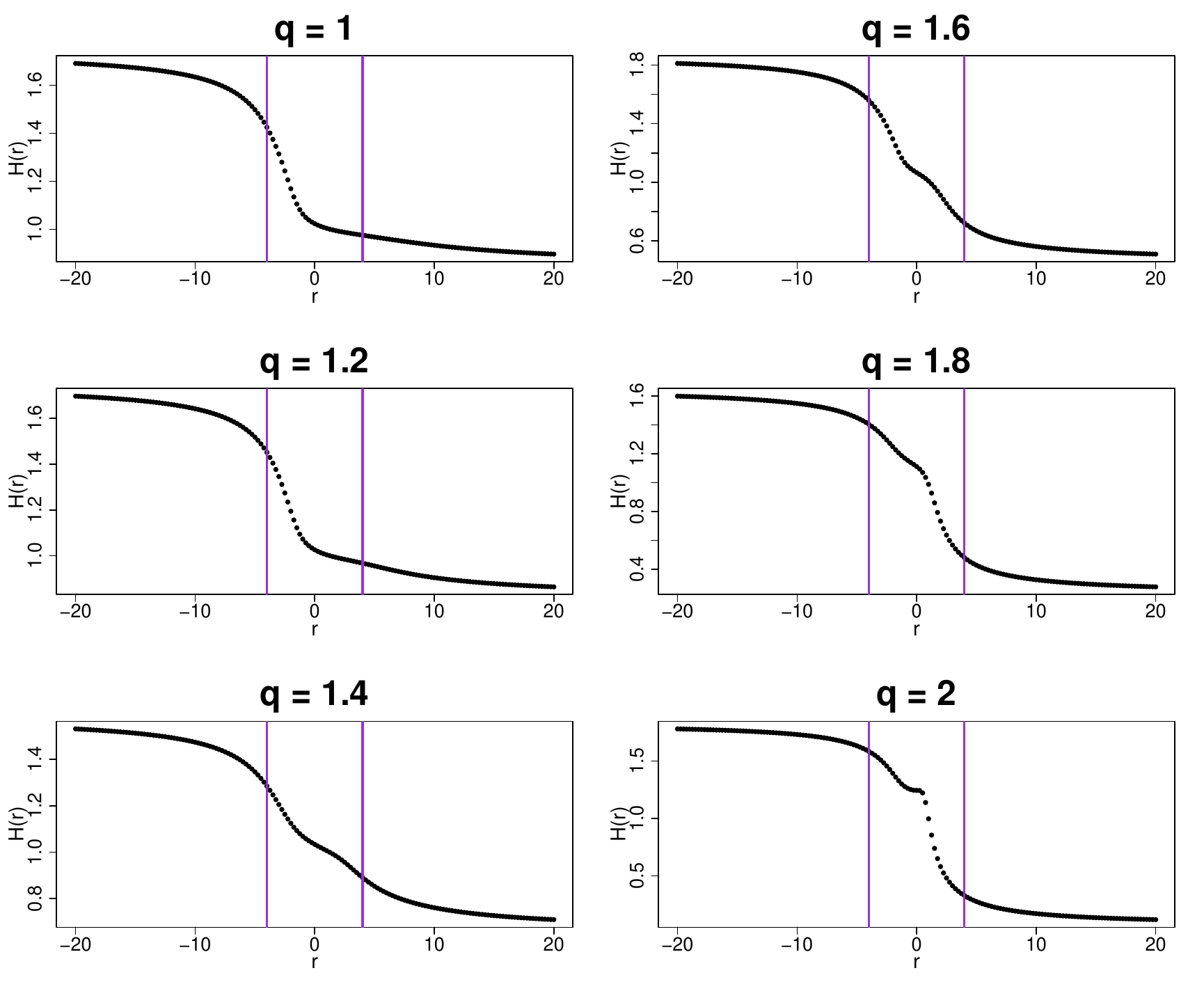}}
\caption{(a) Fluctuation functions $F_r(s)$ and (b) the generalized Hurst exponents $h(r)$ calculated by using MFDFA from $q$-Gaussian time series with $1 \leqslant q \leqslant 2$ and their temporal organization inherited from a dyadic log-gamma cascade. Vertical lines denote the lower and upper boundaries of (a) the range of scales $s$ used to calculate $h(r)$ and (b) the range $-4 \leqslant r \leqslant 4$ used to calculate singularity spectra $f(\alpha)$ shown in Fig.~\ref{fig::transformed.log-gamma.heavy-tails.falpha}. The results have been averaged over 10 independent realizations of the $q$-Gaussian time series.}
\label{fig::transformed.log-gamma.heavy-tails.Frh}
\end{figure}

\begin{figure}
\centering
\includegraphics[width=1\textwidth]{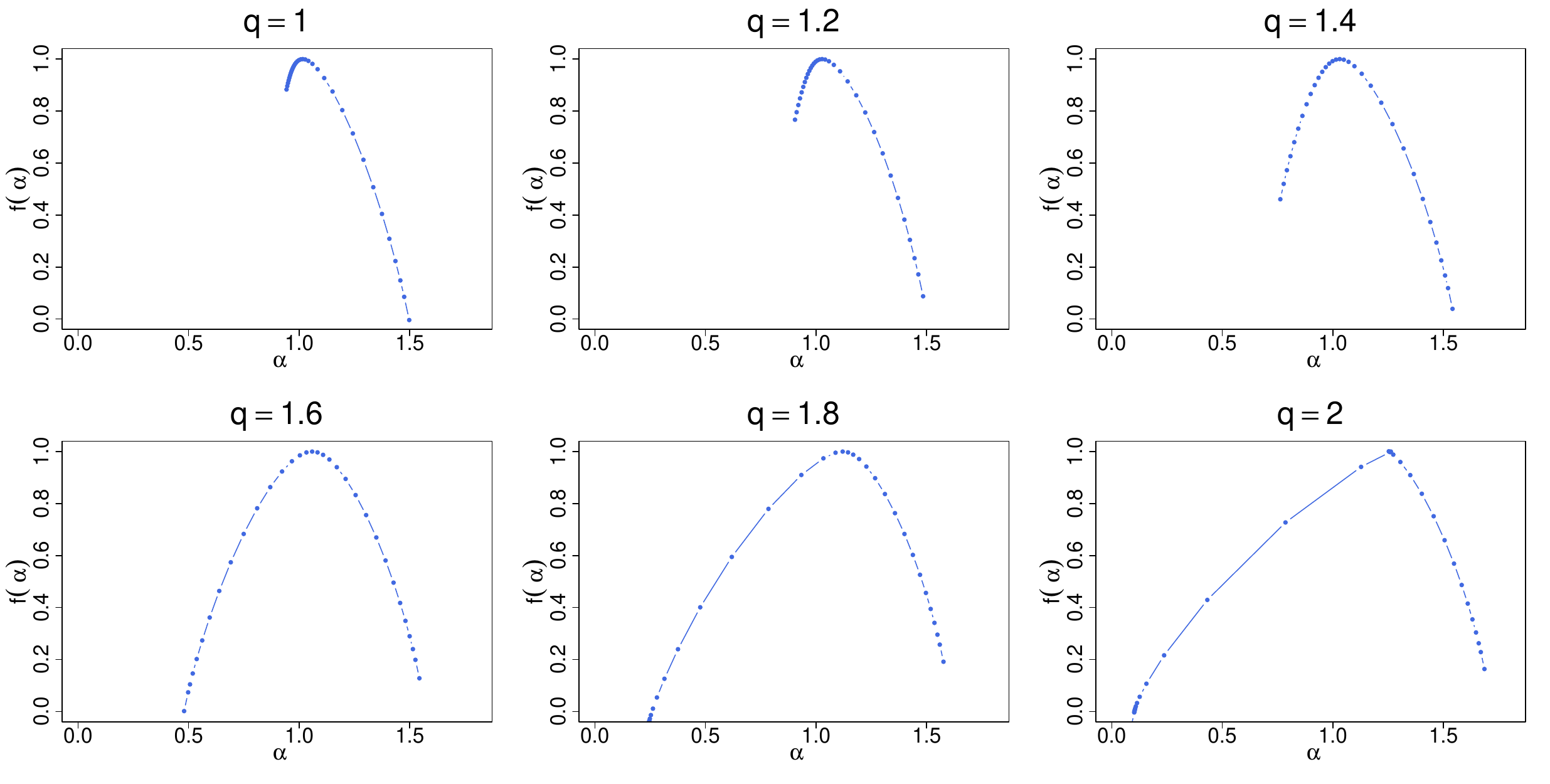}
\caption{Singularity spectra $f(\alpha)$ for the same time series as in Fig.~\ref{fig::transformed.log-gamma.heavy-tails.Frh}. The results have been averaged over 100 independent realizations of the $q$-Gaussian time series.}
\label{fig::transformed.log-gamma.heavy-tails.falpha}
\end{figure}

\begin{figure}
\includegraphics[width=0.9\textwidth]{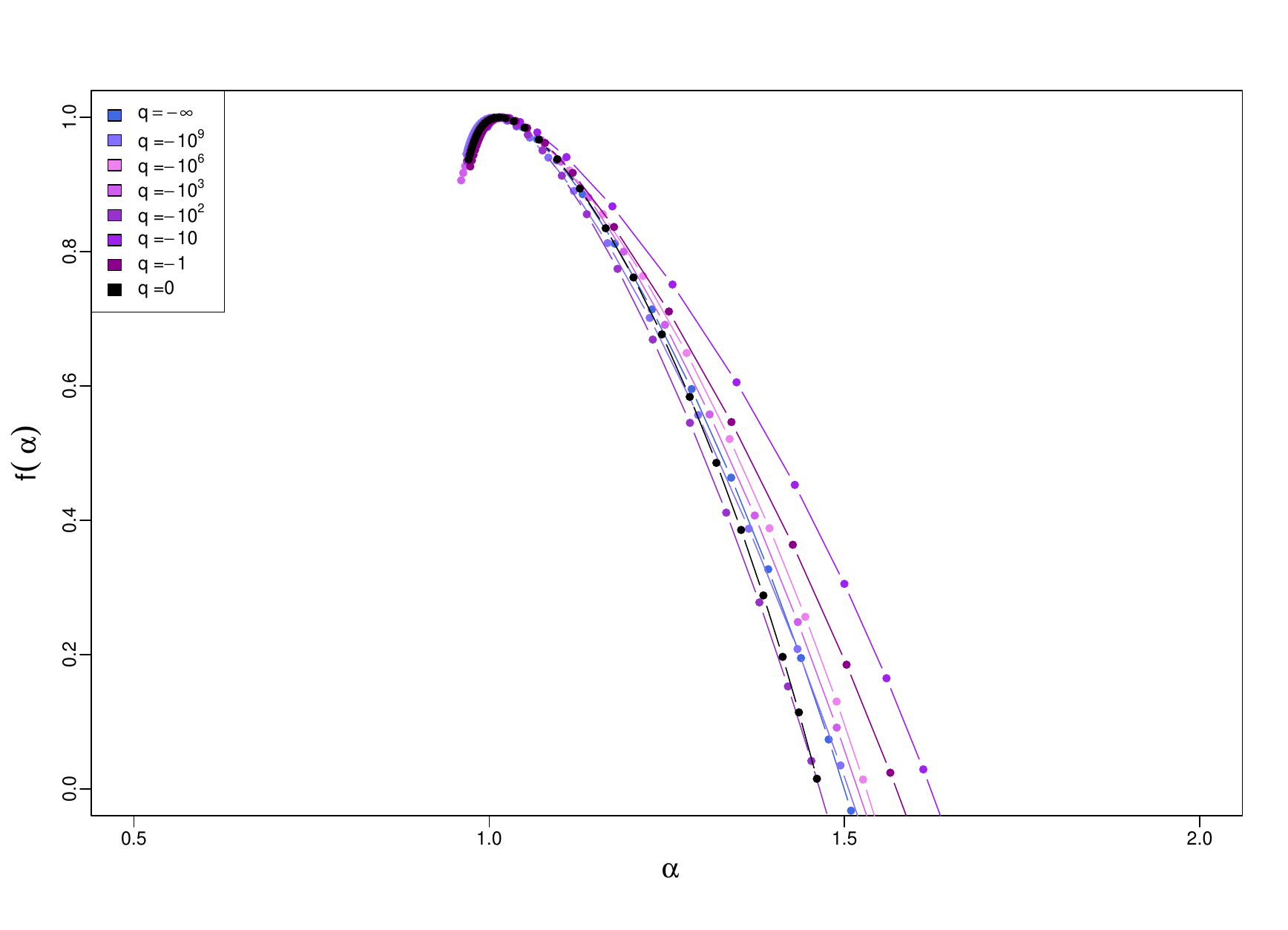}
\caption{Singularity spectra $f(\alpha)$ for time series with $q$-Gaussian PDFs on a compact support, where $-\infty < q < 1$, and with temporal organization inherited from a dyadic log-gamma cascade. The results have been averaged over 10 independent realizations of the $q$-Gaussian time series.}
\label{fig::transformed.log-gamma.no-tail.falpha}
\end{figure}

\begin{figure}
\includegraphics[width=0.9\textwidth]{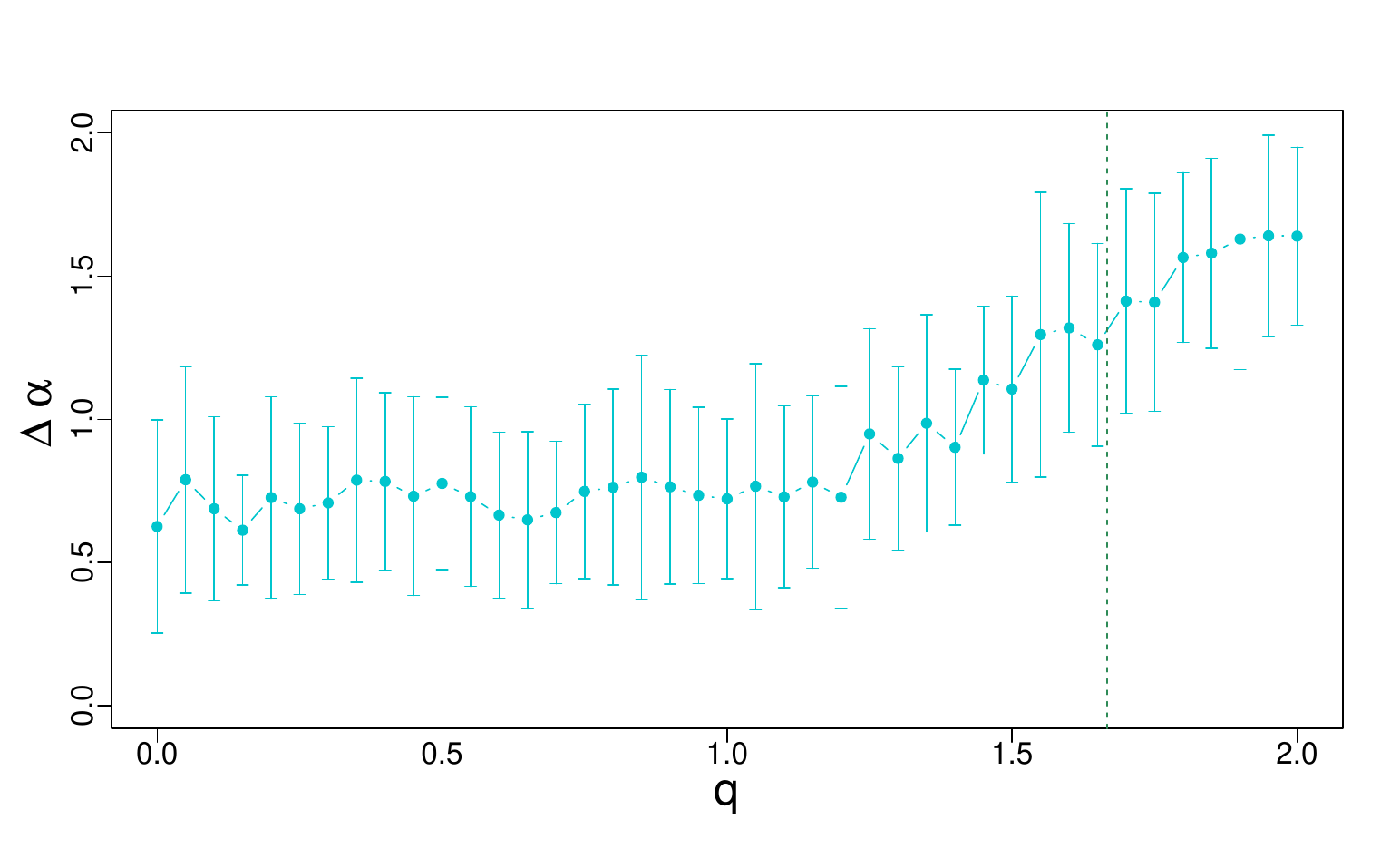}
\caption{Width $\Delta\alpha$ of the singularity spectra $f(\alpha)$ calculated for $q$-Gaussian time series of length of $N=10^5$ data points with $0 \leqslant q \leqslant 2$ and their temporal organization inherited from a dyadic log-gamma cascade. Vertical line at $q=5/3$ separates the Gaussian and L\'evy-Gnedenko basins. Error bars denote standard deviation of $\Delta\alpha$ calculated from 20 independent realizations of the $q$-Gaussian time series.}
\label{fig::transformed.log-gamma.falpha.width}
\end{figure}

Another type of temporal organization is provided by the log-gamma cascades. As before, their original PDFs have been replaced by the $q$-Gaussian ones. Fig.~\ref{fig::transformed.log-gamma.heavy-tails.Frh} shows both the fluctuation functions $F_r(s)$ and the generalized Hurst exponents $h(r)$, the latter being derived from the former. By examining the plots for different values of $q$, one can infer that the range of scales associated with acceptable scaling behavior is narrower, and its lower bound is shifted toward the larger scales compared to the case of the binomial and log-normal cascades. However, the singularity spectra calculated from $h(r)$ do not significantly differ from their counterparts for these other cascades - see Fig,~\ref{fig::transformed.log-gamma.heavy-tails.falpha}. The spectra representing the PDFs with a compact support show a greater similarity to their counterparts for the binomial cascades and a smaller one to the spectra for the log-normal cascades as Fig.~\ref{fig::transformed.log-gamma.no-tail.falpha} shows. Regarding the width $\Delta\alpha$ as a function of $q$, Fig.~\ref{fig::transformed.log-gamma.falpha.width} shows a difference in relation to the case of the log-normal cascades. While there are still two regimes that can be distinguished and the cross-over between them is located, roughly, at the same $q$, the constant-width regime is associated with a larger value of the average width: $\Delta\alpha \approx 0.7$ with its oscillations reaching 0.6 and 0.85. The between-samples variability of the width is also larger than in the previous case. In the second regime, one observes a trace of saturation of the average $\Delta\alpha$ near 1.6 for $q \geqslant 1.8$, which was hardly visible in the log-normal case.


\subsubsection{$q$-Gaussian with a log-Poisson-cascade organization}

\begin{figure}
\centering
\subfloat[]{\includegraphics[trim={0 20 0 0},clip,width=0.85\textwidth]{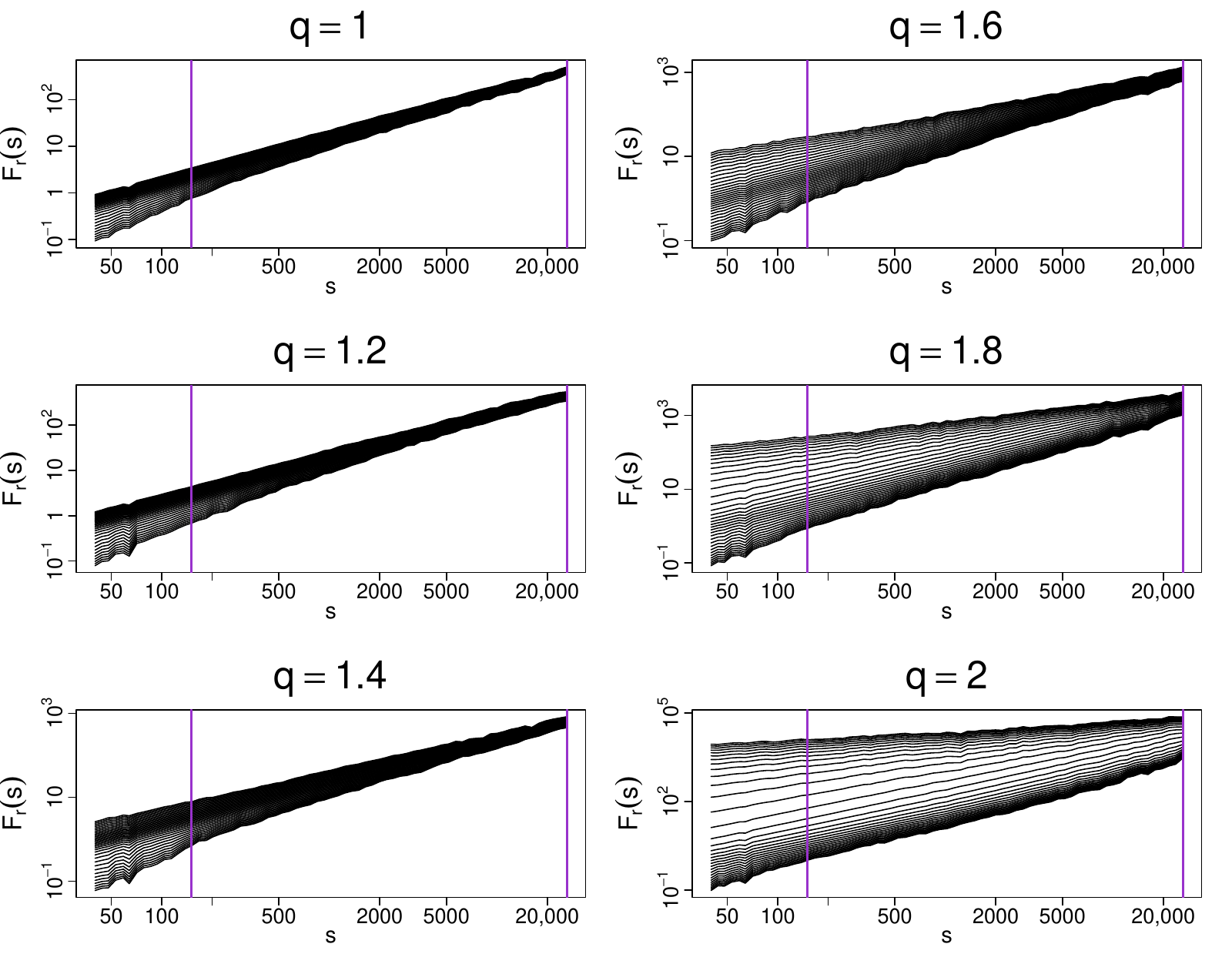}}

\vspace{0.2cm}
\subfloat[]{\includegraphics[trim={0 20 0 0},clip,width=0.85\textwidth]{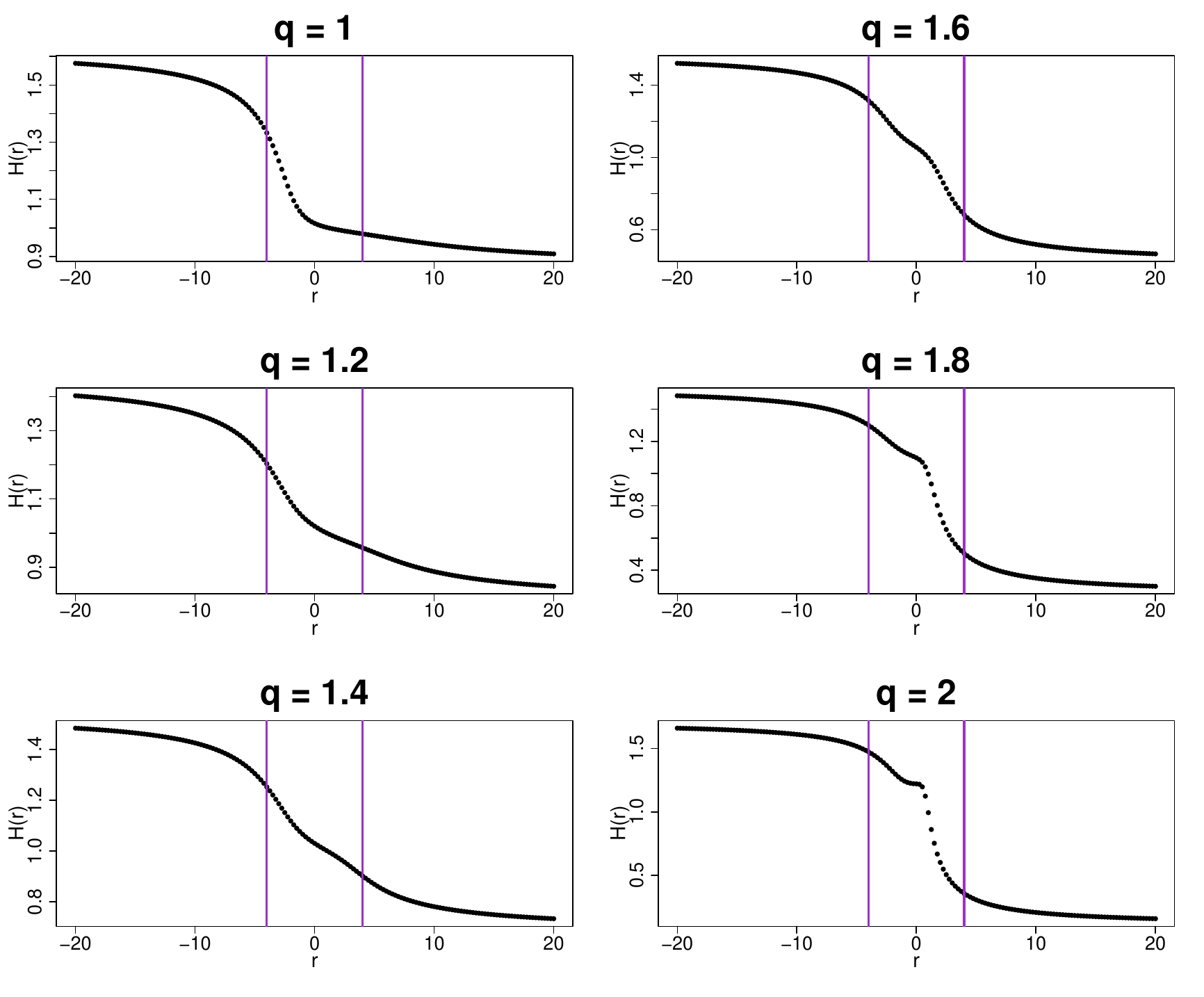}}
\caption{(a) Fluctuation functions $F_r(s)$ and (b) the generalized Hurst exponents $h(r)$ calculated by using MFDFA from $q$-Gaussian time series with $1 \leqslant q \leqslant 2$ and their temporal organization inherited from a dyadic log-Poisson cascade. Vertical lines denote the lower and upper boundaries of (a) the range of scales $s$ used to calculate $h(r)$ and (b) the range $-4 \leqslant r \leqslant 4$ used to calculate singularity spectra $f(\alpha)$ shown in Fig.~\ref{fig::transformed.log-poisson.heavy-tails.falpha}. The results have been averaged over 10 independent realizations of the $q$-Gaussian time series.}
\label{fig::transformed.log-poisson.heavy-tails.Frh}
\end{figure}

In the final part of this study, a set of time series with a temporal organization matching the log-Poisson cascades is considered. If one compares the $F_r(s)$ functions depicted in Fig.~\ref{fig::transformed.log-poisson.heavy-tails.Frh}(a) with the ones analyzed before, a qualitative difference can be found for $q=1.6$. Unlike the binomial, log-normal, and log-gamma cases, here a cross-over point between two scaling regimes with different dispersion strength of the lines representing $F_r(s)$ do exist. Both the regimes can be interpreted as multifractal but the small-scale one is richer. This situation resembles to some extent the one observed for the no-correlation case (see Fig.~\ref{fig::qgaussian.uncorrelated.fluctuation-functions}) but with a more stable scaling over a longer range of scales in the present case. This means that one may expect that, if longer time series were considered, the relative lengths of the two scaling-regime intervals could be changed in favor of the large-scale regime. In spite of some resemblance, there is an actual discrepancy between the present case and the no-correlation one: there is no doubt that both regimes are multifractal now, so one cannot blame the finite-size effects for producing an apparent multiscaling in an truly monofractal data set. It seems that in the log-Poisson cascade, correlations can produce two multifractal regimes if the tails are sufficiently heavy yet still belonging to the Gaussian basin. For larger $q$, such duality is no longer detected.


\begin{figure}
\centering
\includegraphics[width=1\textwidth]{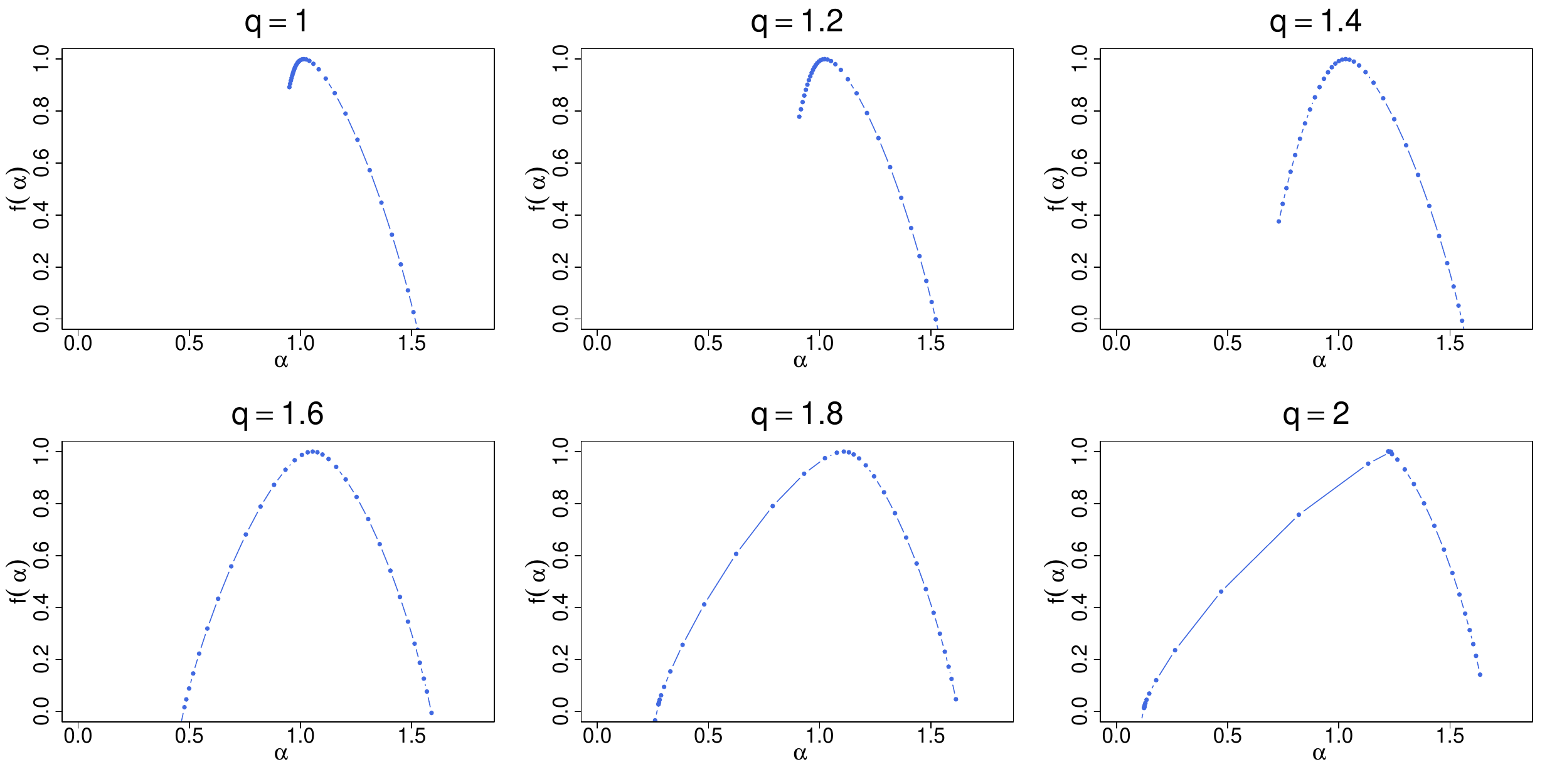}
\caption{Singularity spectra $f(\alpha)$ for the same time series as in Fig.~\ref{fig::transformed.log-poisson.heavy-tails.Frh}. The results have been averaged over 100 independent realizations of the $q$-Gaussian time series.}
\label{fig::transformed.log-poisson.heavy-tails.falpha}
\end{figure}  

\begin{figure}
\includegraphics[width=0.9\textwidth]{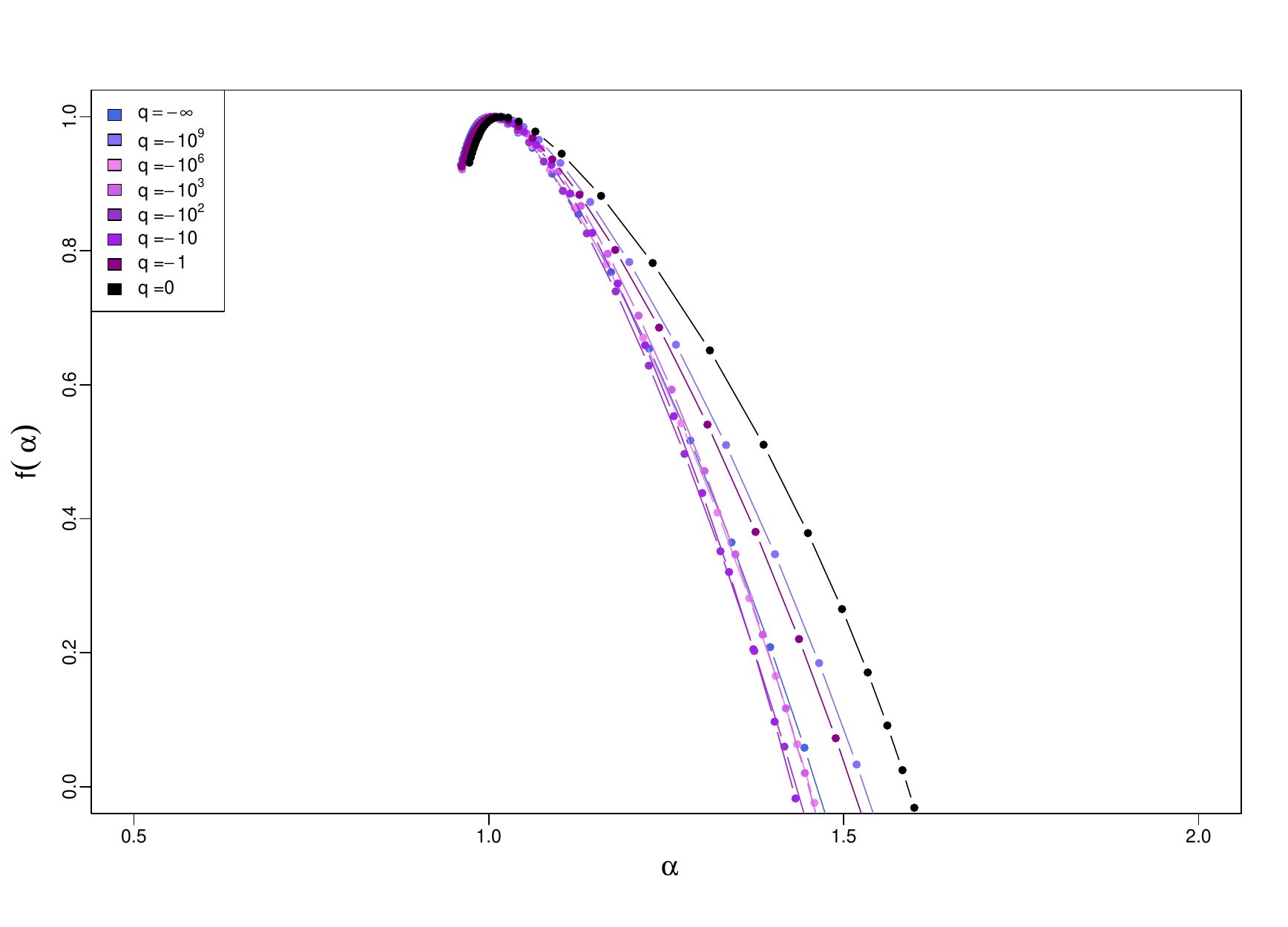}
\caption{Singularity spectra $f(\alpha)$ for time series with $q$-Gaussian PDFs on a compact support, where $-\infty < q < 1$, and with temporal organization inherited from a dyadic log-Poisson cascade. The results have been averaged over 10 independent realizations of the $q$-Gaussian time series.}
\label{fig::transformed.log-poisson.no-tail.falpha}
\end{figure}

\begin{figure}
\includegraphics[width=0.9\textwidth]{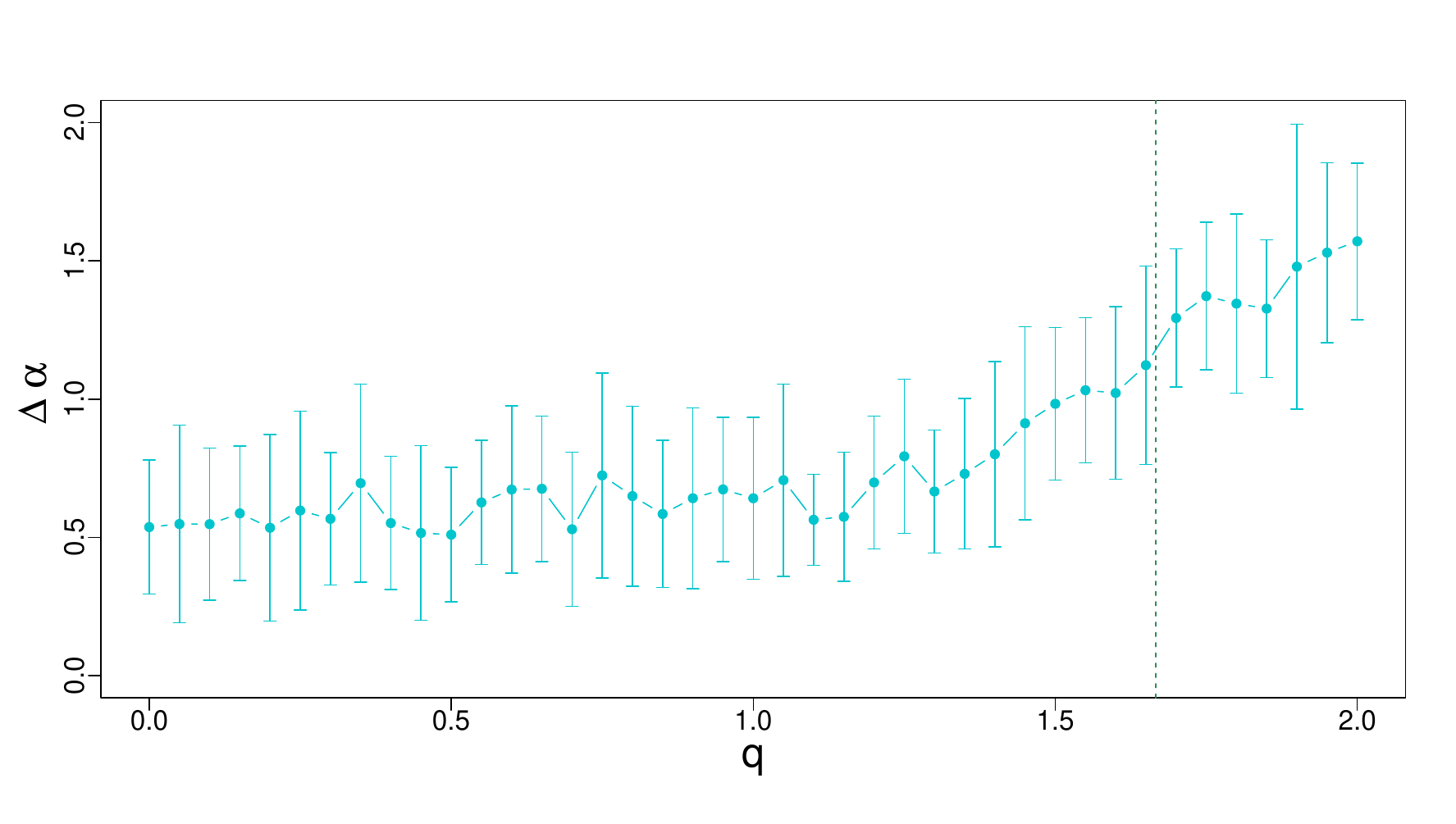}
\caption{Width $\Delta\alpha$ of the singularity spectra $f(\alpha)$ calculated for $q$-Gaussian time series of length of $N=10^5$ data points with $0 \leqslant q \leqslant 2$ and their temporal organization inherited from a dyadic log-Poisson cascade. Vertical line at $q=5/3$ separates the Gaussian and L\'evy-Gnedenko basins. Error bars denote standard deviation of $\Delta\alpha$ calculated from 20 independent realizations of the $q$-Gaussian time series.}
\label{fig::transformed.log-poisson.falpha.width}
\end{figure}

Fig.~\ref{fig::transformed.log-poisson.heavy-tails.Frh}(b) shows that the generalized Hurst exponents do not differ from their counterparts for the other considered cascade types. Also, the spectra $f(\alpha)$ for the PDFs with non-compact support ($1 \leqslant q \leqslant 2$) look close to their counterparts shown before $-$ see Fig.~\ref{fig::transformed.log-poisson.heavy-tails.falpha}. If the PDFs have a compact support, the singularity spectra exhibit a transient behavior in the quantitative terms, located between the binomial and log-normal cases as Fig.~\ref{fig::transformed.log-poisson.no-tail.falpha} documents. However, qualitatively there is no significant difference between this and the cases discussed before. The width of the singularity spectra shown in Fig.~\ref{fig::transformed.log-poisson.falpha.width} assumes transient characteristics as well: the average $\Delta\alpha$ varies between 0.5 and 0.6 within its stationary regime for $q \leqslant 1.2$ with the standard deviation of intermediate size. In the second regime for $q \geqslant 1.3$, the width increases almost linearly with $q$ and no saturation effect can be seen, which is closer to the result for the log-normal cascades (Fig.~\ref{fig::transformed.log-normal.falpha.width}) than to the result for the log-gamma cascades (Fig.~\ref{fig::transformed.log-gamma.falpha.width}).


\section{Conclusions}
\label{sect::conclusions}

In the present study, the impact of the temporal correlations on the multifractal structure of time series was investigated together with the multifractal-building ``cooperation'' of these correlations and the PDF structure in the presence of heavy tails. Based on artificial time series with $q$-Gaussian PDF, it has been shown by using MFDFA that only two results are possible in the no-correlation case: a monofractal structure if the analyzed time series belong to the Gaussian basin of attraction (in the sense of the central limit theorem) and a bifractal structure if the time series belong to the L\'evy-Gnedenko basin. In both situations, a broadening of the singularity spectra $f(\alpha)$ is expected, which can lead to a spurious detection of multifractality that can be especially misleading in the bifractal case, when the singularity spectrum even naturally extends over a range of $\alpha$.

In the second part of the study, temporal correlations $-$ both linear and nonlinear $-$ were introduced to match the temporal organization of multiplicative cascades of different types. Through the rank-ordering density transformation, the original PDFs of the cascades were replaced by $q$-Gaussians with the temporal organization of the original cascades preserved. In this manner, a set of time series with temporal correlations and the $q$-Gaussian PDFs controlled by the Tsallis parameter $q$ were prepared for the multifractal analysis. Based on the same ranking-based temporal structure of the data, by varying $q$ from the one matching a uniform-PDF on a compact support ($q \to -\infty$), via the standard Gaussian case ($q=1$), to the heavy-tail PDFs on a non-compact support ($1 \leqslant q \leqslant 2$), it has been shown that the very existence of (nonlinear) temporal correlations leads to the emergence of the multifractal properties quantified in terms of the width of $f(\alpha)$. This is the best evident in the case of the standard Gaussian PDF, which corresponds to an obvious monofractal structure in the absence of correlations, but develops clear multifractality if the cascade-like correlations are added. The singularity spectra are strongly right-side asymmetric here, which points out to the small fluctuations as a source of multifractality in this case. If the long-range correlations are additionally accompanied by heavy tails, the width of the spectrum gradually grows beyond that observed for the Gaussian PDF. It has also been shown that the width $\Delta\alpha$ increases even if the baseline (no-correlation) case has already been bifractal. In this situation, the spectrum can broaden beyond the width defined by the two proper points (0,0) and ($1/\alpha_{\rm L},1$) and become truly multifractal.

By applying various types of multiplicative cascades $-$ ranging from a deterministic binomial cascade to stochastic canonical cascades with log-normal, log-gamma, and log-Poisson multipliers $-$ it has been demonstrated that the specific form of temporal organization in time series is only secondary to the very existence of long-range nonlinear correlations. The presented results add arguments to the already proven and documented fact in literature that a genuine multifractal structure of time series is possible exclusively under the influence of temporal correlations~\cite{DrozdzS-2009a,KwapienJ-2023a}. However, if a time series is already multifractal because of the presence of long-range correlations, then the particular heavy-tail shape of the PDF can amplify the multifractal structure and increase complexity of such a time series, indeed. The progress with respect to refs~\cite{DrozdzS-2009a,KwapienJ-2023a} is now such that a quantitative estimate of the influence of the fluctuation distribution on the width of the multifractal spectrum $f(\alpha)$ is provided. As a side-result of the present study, it has been documented for the first time in literature that time series whose PDF has a compact support can develop a multifractal structure in the same way as the Gaussian time series do. It occurs that the main factor responsible for this property is the attraction basin a given PDF belongs to, while the exact shape of the PDF is less important in this context. Therefore, when looking for an answer to the frequently asked question about the share of the fluctuation distribution in the total multifractal spectrum of a time series, the only scientifically valid approach is to estimate the surplus in relation to the series with the same temporal correlations but the fluctuation distribution reduced to the Gaussian distribution. The above-presented procedure of projecting onto $q$-Gaussian distributions, with $q=1$ as an appropriate reference for this particular purpose, can be successfully used.  Based on previous results showing that wavelet methods for describing multifractality are also effective when applied to cascades~\cite{DrozdzS-2009a,OswiecimkaP-2006a} it can be expected that the disentanglement procedure presented here can also perform well in combination with wavelet methods. Finally, this procedure opens a perspective on the analysis of empirical structures and time series. In particular, it will be the topic of a separate, this time empirically oriented paper by the current authors.

\authorcontributions{Conceptualization, R.K., S.D., J.K., T.S., and M.W..; methodology, R.K., S.D., J.K., T.S., and M.W.; software, R.K.; validation, R.K., S.D., J.K., and T.S.; formal analysis, R.K., S.D., J.K., and T.S.; investigation, R.K. and T.S.; resources, R.K. and T.S.; data curation, R.K.; writing---original draft preparation, S.D., R.K. and J.K.; writing---review and editing, R.K., S.D., J.K., T.S., and M.W.; visualization, R.K.; supervision, S.D.; project administration, S.D. and M.W.; funding acquisition, M.W. All authors have read and agreed to the published version of the manuscript.}

\funding{This work was partially supported by the Polish National Science Center grant no. 2023/07/X/ST6/01569.}

\institutionalreview{Not applicable.}

\informedconsent{Not applicable.}

\dataavailability{The time series generated for this study are available from the authors on reasonable request.} 

\acknowledgments{M.W. thanks the Adapt Centre, School of Computing, Dublin City University, Dublin, Ireland and, in particular, its head, prof. Martin Crane, for hospitality during his stay.}

\conflictsofinterest{The authors declare no conflicts of interest.} 

\abbreviations{Abbreviations}{
The following abbreviations are used in this manuscript:\\

\noindent 
\begin{tabular}{@{}ll}
MFDFA & Multifractal detrended fluctuation analysis\\
PDF & Probability distribution function
\end{tabular}
}

\appendixtitles{no}
\appendixstart
\appendix
\section[\appendixname~\thesection]{}

Proof of Eq.~(\ref{eq::qgaussian.minus-infinity}). First, a few basic properties of the gamma function can be noted:
\begin{equation}
\Gamma(z+1) = z\Gamma(z) \qquad \Gamma\left(\frac{1}{2}\right)=\sqrt{\pi} \qquad \Gamma(1) = 0! =1
\end{equation}
for $z\in\mathbb{C}$ with $\mathfrak{Re}(z)>0$ being continuous in $\mathbb{R}$ on the interval $(0,\infty)$. Thus, given a number $c\in\mathbb{R}$ and a function $f:\mathbb{R}\to\mathbb{R}$ with the limit $d:=\lim_{x\to c} f(x)$, such that $\Gamma$ is continuous at $d$, one can write:
\begin{equation}
\lim_{x\to c} \Gamma(f(x)) = \Gamma\left(\lim_{x\to c}f(x)\right) = \Gamma(d).
\end{equation}
It is also known that, as $q \to -\infty$, then $1-q \to \infty$ and $1/(1-q)\to 0$. With that in mind, one can quickly calculate the following limits:
\begin{gather*}
\lim_{q\to-\infty} \sqrt{\frac{3-q}{1-q}}= \lim_{q\to-\infty} \sqrt{1+\frac{2}{1-q}} = 1 \\
\lim_{q\to-\infty} \frac{3-q}{2(1-q)} = \frac{1}{2}\lim_{q\to-\infty} \left(1+\frac{2}{1-q}\right) = \frac{1}{2}.
\end{gather*}
Therefore,
\begin{align*}
\lim_{q\to-\infty} \frac{1}{\sqrt{3-q}C_q} =\lim_{q\to-\infty} \frac{(3-q)\sqrt{1-q}\Gamma\left(\frac{3-q}{2(1-q)}\right)}{2\sqrt{\pi}\sqrt{3-q}\Gamma\left(\frac{1}{1-q}\right)}&=\lim_{q\to-\infty}\frac{\sqrt{\frac{3-q}{1-q}}\Gamma\left(\frac{3-q}{2(1-q)}\right)}{2\sqrt{\pi}\frac{1}{1-q}\Gamma\left(\frac{1}{1-q}\right)} =\\
= \frac{\lim_{q\to-\infty}\sqrt{\frac{3-q}{1-q}} \cdot\lim_{q\to-\infty}\Gamma\left(\frac{3-q}{2(1-q)}\right) }{2\sqrt{\pi}\lim_{q\to-\infty}\frac{1}{1-q}\Gamma\left(\frac{1}{1-q}\right)} &=
\frac{1\cdot \Gamma\left(\lim_{q\to-\infty}  \frac{3-q}{2(1-q)}\right)}{2\sqrt{\pi}\lim_{q\to-\infty}\Gamma\left(\frac{1}{1-q}+1\right)} =\\
= \frac{\Gamma\left(\frac{1}{2}\right)}{2\sqrt{\pi}\Gamma\left(\lim_{q\to-\infty}\frac{1}{1-q}+1\right)} &= \frac{\sqrt{\pi}}{2\sqrt{\pi}\Gamma(1)} = \frac{1}{2}.
\end{align*}
Now it remains to show that $\lim_{q\to-\infty} \left(1-\frac{1-q}{3-q}x^2\right)^{\frac{1}{1-q}} =1 $. Let one define
\begin{equation}
f(q) = 1-\frac{1-q}{3-q}x^2=1-x^2+\frac{2x^2}{3-q}\qquad g(q) = \frac{1}{1-q}
\end{equation}
and calculate
\begin{equation}
f'(q) = \frac{2x^2}{(3-q)^2} \qquad g'(x) = \frac{1}{(1-q)^2}.
\end{equation}
It is important to notice that, if $x\in[-1,1]$ and $q<3$, one has $f(q)>0$. Then
\begin{align*}
\lim_{q\to-\infty}f(q) &=  \lim_{q\to-\infty} 1-x^2+\frac{2x^2}{3-q}=1-x^2\\
\lim_{q\to-\infty} g(q) &= 0.
\end{align*}
Having that calculated, one can start with
\begin{equation}
\lim_{q\to-\infty}\left(1-\frac{1-q}{3-q}x^2\right)^{\frac{1}{1-q}} = \lim_{q\to-\infty} f(q)^{g(q)} = \lim_{q\to-\infty} \exp\left(\ln(f(q)) g(q)\right) = 
\end{equation}
and, given that $e^x$ is continuous on entire $\mathbb{R}$, one can continue to
\begin{equation}
=\exp\left(\lim_{q\to-\infty}\ln(f(q)) g(q)\right).
\end{equation}
Now, let one assume that $x\neq\pm1$. Then, given that natural logarithm is continuous in $1-x^2>0$ as it is continuous on $(0,\infty)$, it follows that
\begin{equation}
\lim_{q\to-\infty}\ln(f(q)) = \ln\left(\lim_{q\to-\infty} f(q)\right) = \ln(1-x^2)\\
\end{equation}
and, therefore,
\begin{equation}
\lim_{q\to-\infty}\left(1-\frac{1-q}{3-q}x^2\right)^{\frac{1}{1-q}} = \exp\left(\lim_{q\to-\infty}\ln(f(q)) g(q)\right) = \exp(\ln(1-x^2)\cdot 0) = 1.
\end{equation}
When $x=\pm 1$ or $x^2=1$, one can continue with
\begin{equation}
\exp\left(\lim_{q\to-\infty}\ln(f(q)) g(q)\right) = \exp \left(\lim_{q\to-\infty}\frac{\ln(f(q))}{1/g(q)} \right) =
\end{equation}
and then, as $\lim_{q\to-\infty} f(q) = 0^+$, which means $\lim_{q\to-\infty} \ln(f(q)) = -\infty$ as well as $\lim_{q\to-\infty} 1/g(q) = \infty$, L'Hospital's rule allows one to continue the calculation:
\begin{align*}
=\exp\left( \lim_{q\to-\infty} \frac{(\ln(f(q)))'}{(1/g(q))'} \right) &= \exp\left( \lim_{q\to-\infty} \frac{f'(q)/f(q)}{(1-q)'} \right) =\\
=\exp\left(\lim_{q\to-\infty} \frac{\frac{2}{(3-q)^2}}{(-1)\cdot \left(1-1+\frac{2}{3-q}\right)} \right) &= \exp\left(\lim_{q\to-\infty}\frac{1}{q-3}\right) = \exp(0)=1.
\end{align*}
In summary, both limits:
\begin{equation}
\lim_{q\to-\infty} \frac{1}{\sqrt{3-q}C_q} = \frac{1}{2} \qquad \lim_{q\to-\infty} \left(1-\frac{1-q}{3-q}x^2\right)^{\frac{1}{1-q}} =1
\end{equation}
have been proven, which provides one with the desired equation.\qed

\reftitle{References}

\bibliography{bibliography_file}

\PublishersNote{}

\end{document}